\documentclass[smallextended]{svjour3}       
\usepackage{graphicx}

\usepackage{booktabs} 
\usepackage[noorphans,vskip=0.5ex]{quoting}
\usepackage{tabularx, makecell}
\usepackage{moreverb}
\usepackage{color}
\usepackage{multirow}
\usepackage{listings}
\usepackage{color}
\usepackage{caption}
\usepackage{float}
\usepackage{subcaption}
\definecolor{dkgreen}{rgb}{0,0.6,0}
\definecolor{gray}{rgb}{0.5,0.5,0.5}
\definecolor{mauve}{rgb}{0.58,0,0.82}
\lstdefinestyle{customc}{
  breaklines=true,
  xleftmargin=\parindent,
  language=Java,
  showstringspaces=false,
  keywordstyle=\bfseries\color{green!40!black},
  commentstyle=\itshape\color{purple!40!black},
  identifierstyle=\color{blue},
  stringstyle=\color{orange},
}

\titlerunning{Explicit Programming Strategies}
\authorrunning{LaToza, Arab, Loksa, and Ko}

\begin{document}
%
\title{Explicit Programming Strategies}
%
%
%
%

\author{Thomas D. LaToza \and
        Maryam Arab \and
        Dastyni Loksa \and 
        Amy J. Ko
}        

\institute{ Thomas D. LaToza and Maryam Arab  \at
    George Mason University, Fairfax, VA, USA \\
    \email{\{tlatoza, marab\}@gmu.edu } \\
        \and
    Dastyni Loksa and Amy J. Ko \\
    University of Washington, Seattle, WA, USA \\
    \email{\{ajko, dloksa\}@uw.edu} \\
}

\markboth{In review}%
{Explicit Programming Strategies}
\maketitle

\begin{abstract}
  Software developers solve a diverse and wide range of problems. While software engineering research often focuses on tools to support this problem solving, the strategies that developers use to solve problems are at least as important. In this paper, we offer a novel approach for enabling developers to follow explicit programming strategies that describe how an expert tackles a common programming problem. We define explicit programming strategies, grounding our definition in prior work both within software engineering and in other professions which have adopted more explicit procedures for problem solving. We then present a novel notation called Roboto and a novel strategy tracker tool that explicitly represent programming strategies and frame executing strategies as a collaborative effort between human abilities to make decisions and computer abilities to structure process and persist information. In a formative evaluation, 28 software developers of varying expertise completed a design task and a debugging task. We found that, compared to developers who are free to choose their own strategies, developers given explicit strategies experienced their work as more organized, systematic, and predictable, but also more constrained. Developers using explicit strategies were objectively more successful at the design and debugging tasks. We discuss the implications of Roboto and these findings, envisioning a thriving ecosystem of explicit strategies that accelerate and improve developers' programming problem solving.
  
\end{abstract}

\section{Introduction}
\label{sec:introduction}

Programmer productivity has been a major focus of software engineering research for decades. The field has studied tools to make developers more productive (e.g.,\cite{kersten2006using}), it has long investigated measures of individual developer productivity (e.g.\cite{sackman1968exploratory}), it has tried to predict productivity from numerous factors (e.g., \cite{dieste2017empirical}), and it has contextualized productivity among the larger set of skills that developers must have to be great engineers (e.g., \cite{li2015makes}). At the heart of all of these efforts is the goal of understanding what factors make developers productive.

Much of this research has focused on specific skills. For example, researchers have captured low-level programming knowledge as part of learning technologies \cite{anderson1989skill}. A long history of work has theorized about program comprehension skills, describing the bottom up, top down, and opportunistic strategies that developers use at a high level \cite{roehm2012professional,von1995program}. Some have theorized about debugging strategies, characterizing them as iterative convergent processes \cite{gilmore1991models}. More broadly, research on program comprehension suggests that ``experts seem to acquire a collection of strategies for performing programming tasks, and these may determine success more than does the programmer's available knowledge" \cite{gilmore1990expert}.

However, developers today have few explicit programming strategies which they may apply. An explicit programming strategy is a \textit{human-executable procedure for accomplishing a programming task}. It describes steps in which a human acts or retrieves and interprets information from the world.  
Consider, for example, debugging: there are many abstract descriptions of what developers do when debugging (e.g., forming hypotheses, gathering data) \cite{zeller2009programs}, but few explicit strategies which enumerate steps a developer can follow to reliably localize a fault from a failure. Similarly, consider program design. Work has abstractly characterized it as a process of transforming a programming problem into a program that solves it, but few strategies offer explicit steps by which developers can do this activity. Or consider API selection, which likely has many hard-won strategies in industry, few of which have been written down systematically as a strategy. While programmers perform these software engineering skills daily, we have few explicit strategies for doing these tasks that developers can follow, either to improve their performance, or to learn the skill for the first time. These strategies, reflecting developers processes and skills, therefore remain invisible, mostly residing in the minds of the world's most experienced developers.

In other engineering disciplines, researchers explicitly prescribe procedures in handbooks to train new engineers. Consider, for example, the Civil Engineering Handbook \cite{chen2002civil}. It is a tome, bursting with examples of how to plan and schedule construction, how to process wastewater, and how to design steel structures. Such handbooks not only describe numerous procedures that constitute civil engineering skill, but also provide numerous examples of how to apply these procedures to solve problems in the domain. These books provide engineers with explicit procedures for solving common problems in a field of engineering, and often determine what is taught in engineering classes, what constitutes an accredited curriculum, and what is tested in engineering licensing exams. Handbooks also support the work of expert engineers, providing them reminders, guidance, and evidence-based procedures that present the best practices for solving engineering problems.

Software engineering, in contrast, has no such handbook.  While computer science and software engineering has cataloged algorithms, design patterns, architectural styles and other specific solutions, this knowledge is declarative, rather than procedural in form, describing templates that can be applied in a situation rather than a procedure for taking action.
Yet much of software engineering work constitutes problem solving, where developers seek information and make decisions, such as when localizing a fault, reasoning about the implications of a change, or formulating a design \cite{LaToza2007Fact,Ko2007InfoNeeds,Sillito2008Info}. 
If we had a software engineering handbook encompassing procedural knowledge, novice software engineers might more rapidly learn effective software engineering strategies. Experts might be more productive, following well-tested procedures for solving a range of software engineering problems, rather than taking shortcuts, satisficing, and using organically-developed personal strategies. This formalization of programming problem solving could ultimately result in better software by both accelerating software engineering work and by preventing defects, which often have their root causes in human error \cite{ko2005framework}. 

Before we can create such a handbook for software engineering, there must first be a way of explicitly representing strategies for programming, so that developers may read and follow these strategies to guide their work. Unfortunately, there are many open questions about how to represent strategies for developer use:

\begin{itemize}
    \item How can we \textit{describe} explicit programming strategies?
    \item How can we support developers in following explicit programming strategies?
    \item How do explicit programming strategies help and hinder developers' effectiveness?
\end{itemize}

\begin{figure}[]
\includegraphics[width=0.6\columnwidth, keepaspectratio]{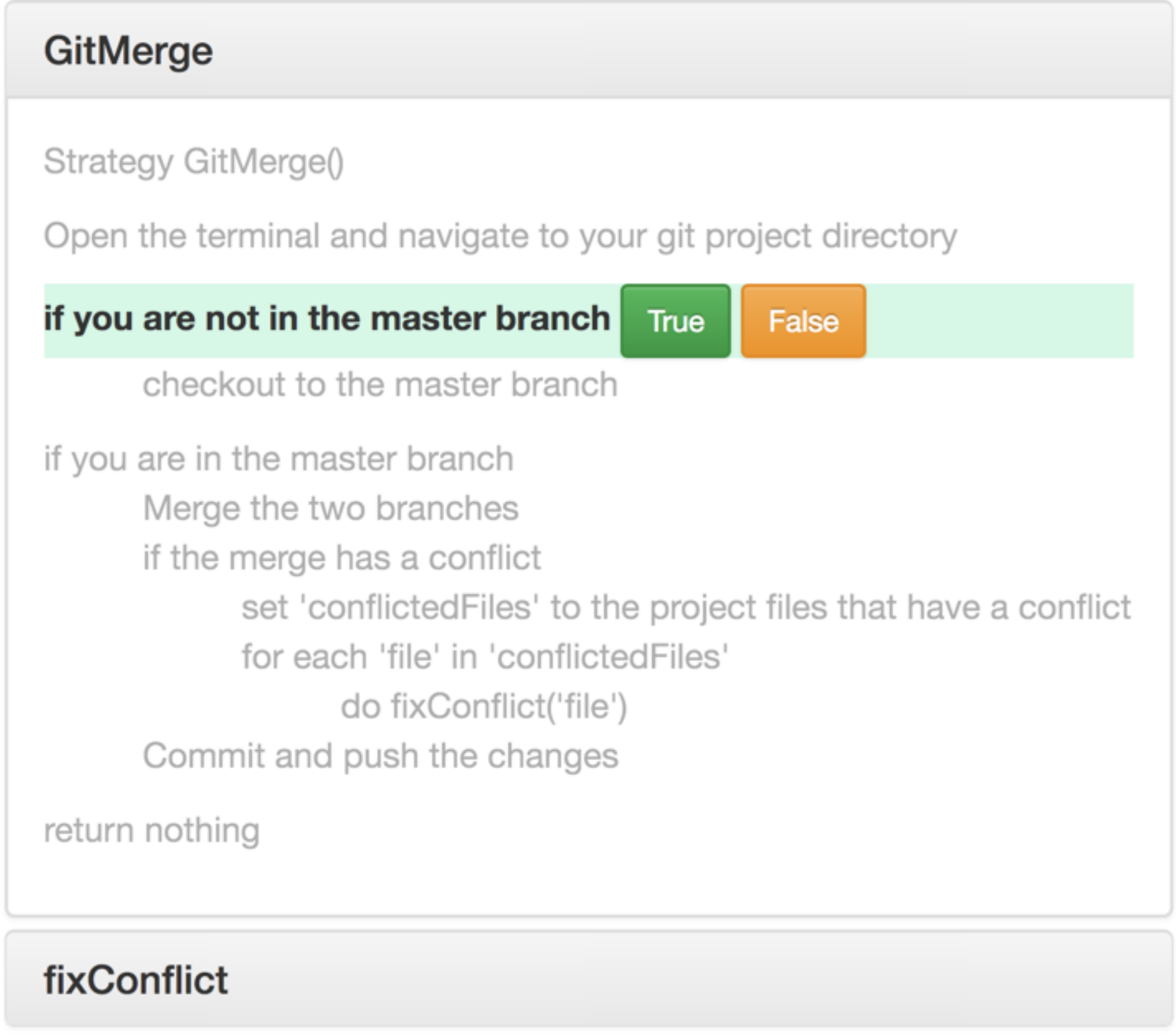} 
\caption{Explicit programming strategies capture a specific sequence of actions for accomplishing a programming task. For example, the above example offers a strategy for merging using Git. To execute this strategy, the strategy tracker lets the computer and developer work together, assigning responsibilities to each. For example, in an if statement, the developer determines if the query is true or false and the computer then advances the program counter to the appropriate next statement.}
\label{figure:Barriers}
\end{figure}

In this paper, we investigate representing programming strategies as semi-formal procedures that involve the productive interaction between a developers' ability to reason and decide and a computer's ability to structure, process, and persist information. We first review related background on strategies, surveying work in other domains as well as software engineering (Section \ref{sec:Background}). We then propose Roboto,
a mixed-initiative strategy description language for explicitly describing programming strategies, and a strategy tracker tool for supporting developers in following these strategies (Figure \ref{figure:Barriers}, Section \ref{sec:ExplicitStrategies}). Through a formative evaluation comparing Roboto strategies to developers' own self-guided strategies, we show that developers were not only more systematic with Roboto strategies but that they also found the explicit support helpful and organizing, and that the explicit strategies led to increased success (Section \ref{sec:Evaluation}). While developers also found explicit strategies constraining, many viewed this as a reasonable tradeoff if the strategies are more effective. We then discuss the implications of this work for software engineering (Section \ref{sec:Discussion}) and conclude (Section \ref{sec:Conclusion}).
\section{Background}
\label{sec:Background}

Academic literature on strategies varies widely in how it defines strategies. For example, in educational psychology, there has long been concern about the lack of a coherent definition: ``[It] appears that ``strategy'' is too broad term and must be defined more specifically for meaningful interpretations to be made.'' \cite{ames1988achievement}. And yet, in psychology, there is a vast literature about the effects of explicit training on strategies, showing the effects of explicit strategies on physics problem solving \cite{ccalicskan2010effects}, the significant role of choice of strategy on task performance \cite{locke1984effect}, and the importance of meta-cognitive strategies in retaining general problem solving among the elderly \cite{labouvie1976cognitive}. In this literature, regardless of how strategies are defined or operationalized, strategy training appears to have strong effects.

It has long been known that humans have limited attention and working memory resources, limiting problem solving performance and leading to human error when limits are exceeded \cite{reason1990}. In response, work in the area of distributed cognition views human problem solving not as computation that occurs exclusively in the head but as computation that is distributed between the human and the artifacts that exist in their environment. Environments in complex domains, such as naval ships, often explicitly offload some responsibility for planning and persistence to the world \cite{Hutchins1995}. Controlled lab studies examining problem solving found that external representations provide memory aids, ease use of information from the environment, structure cognitive behavior, and change the nature of the task \cite{ZHANG1994}. Studies of developers have found that experts offload working memory demands by taking notes during programming tasks \cite{robillard2004effective}.

One form of external representation supporting problem solving is the standard operating procedure (SOP). SOPs formalize complex, error-prone operations, such as landing a plane or repairing a nuclear power plant \cite{wieringa1998procedure}. SOPs offer a series of step-by-step instructions that help people in an organization achieve efficiency, prevent errors, reduce communication, and comply with regulations. Widely deployed in the military, health, and other safety critical settings, SOPs regulate individual, team, and organizational behavior.

In health care, a similar idea of \textit{checklists} has increased in popularity, partly due to Gawande's \textit{Checklist Manifesto} \cite{gawande2010checklist}. This book argued that many problems in medicine are so complex, no individual or team can adequately manage complexity; checklists of actions and states to verify in medical and surgical procedures can manage this complexity by guarding against medical errors and setting a standard of performance. Gawande notes that many providers in health care are resistant to embracing checklists because they take autonomy away from the expert and put it in a document. He argues that the collaborative definition of checklists by teams is key to encouraging experts to embrace explicit strategies, and presents evidence that when they do, medical errors decline and outcomes improve.

Suchman also addressed strategies in her book \textit{Plans and Situated Actions} \cite{suchman1987plans}. In it, she argued that human behavior generally does not emerge from individuals making plans and executing them. Rather, the context in which one is situated, and the rich awareness it affords, make plans more of a resource to draw upon to inform decisions and action. She argues that rather than designing plans and expecting people to follow them, one should design \emph{contexts} in which plans inform and guide behavior, along with other factors, such as the expertise of individuals and teams.

In software engineering, one way to capture expertise is by codifying specific recurring groups of elements, such as in the form of design patterns or architectural styles. For example, Gamma et al. outline a set of design patterns that address common problems in achieving modularity and reuse in object-oriented design \cite{Gamma:1995}. Shaw and Garlan envision a world in which architectural styles are made explicit and shared and outline several common architectural styles \cite{Shaw:1996}. These capture the structure of systems, whereas programming strategies offer complementary expertise, prescribing the structure of developers' work.

Representing strategies through a dedicated description language is an example of a domain-specific language. A domain-specific language offers developers a language specialized for supporting a specific task, offering task-specific notations and constructs that trade generality for fitness for a specific purpose \cite{Mernik2005DSL}. Traditional domain-specific languages focus on specifying computation that is done entirely by the computer. Our work explores how domain-specific languages might be used to describe work that is distributed across both computers and humans. 

Research on programming and software engineering links explicit strategies and productivity. For example, studies have shown that the use of explicit slicing strategies in debugging \cite{francel2001value}, the use of explicit strategies in tracing program execution \cite{xie2018explicit}, and the use of explicit strategies to extract requirements from problem statements \cite{haidry2017identifying}, are either correlated with or cause decreases in task completion time or increases in task performance. Early research on the LISP tutor \cite{anderson1989skill} and software design environments \cite{rist1995program} similarly showed that by defining expert problem solving strategies, and nudging novices to follow those expert strategies through hints and feedback, novices could approach expert performance. Many have also described explicit strategies for debugging; Metzger, for example, describes debugging strategies as a high-level plan for accomplishing a goal through a sequence of physical and cognitive actions \cite{metzger2004debugging} and Zeller presents a series of formal and informal procedures for isolating causes and effects of defects \cite{zeller2009programs}. In the area of software architecture, work has described techniques by which developers should choose between various design alternatives to make design decisions \cite{Bass2012SAIP,Falessi2011ArchDecisions}.

Research on programming has also found that \emph{metacognitive} strategies---helping developers to think more systematically about their thinking and their progress---is both associated with and causes increases in task performance. Early work speculated that ``the strategic elements of programming skill may, in some case, be of greater significance than the knowledge-based components.'' \cite{davies1993models}. Later studies confirmed this for novices, finding that while novices have many diverse strategies while programming \cite{mccartney2007successful}, most of their strategies are either inherently ineffective, executed ineffectively, or interleaved with ineffective strategies \cite{robins2003learning,MurphySIGCSE08,loksa2016role}. To explain these effects, some studies demonstrated that self-regulation strategies, such as monitoring one's work and explicitly evaluating one's progress and use of time, and self-explanation, were associated with greater problem solving success \cite{robillard2004effective,falkner2014identifying}. Other studies showed experimentally that explicit training on these general self-regulation strategies \cite{bielaczyc1995training} or self-regulation strategies specifically related to programming \cite{loksa2016programming}, can cause significant increases in task productivity and programming self-efficacy. Other work has explored the possibility of teaching specific problem solving strategies to novice programmers and the challenges that this brings \cite{Ko2019TeachingStrategies}.

In software engineering, a common way to offer prescriptive problem solving guidance is through a ``practice'' or development methodology. For example, Agile \cite{beck2001manifesto} presents high-level plans and principles for guiding action, offering principles such as ``Continuous attention to technical excellence and good design enhances agility.`` Such general guidelines and principles do not, by themselves, offer step-by-step guidance, but might be used to motivate more detailed strategies describing how they should be carried out. Some prior work has proposed explicit steps for specific activities in software engineering. For example, research on process specification languages attempt to model software development processes to specify steps engineers must follow to achieve a software engineering goal \cite{zamli2001process}. Other work has described steps in a design process, such as laying out a linear design process to specify, implement, test, and review a function \cite{felleisen2001design}. Perhaps the most extensive catalogs exist for refactoring \cite{Fowler1999Refactoring,Kerievsky2004Refactoring}. Test-driven development \cite{beck2003test} offers specific steps developers should use to write tests, then write the minimum amount of code for tests to pass.  
Research has also studied emergent strategies in contexts like pair programming, finding that explicit strategies can evolve organically through negotiation \cite{salinger2013understanding}. Other recent work has investigated mixed-initiative ``lab protocols`` that scientists often write when analyzing data computationally, finding that while lab protocols represent idealized steps, scientists use a range of techniques to expand or limit the semantic interpretation of these protocols \cite{abbott2015programs}. 

While this prior work suggests that explicit strategies and explicit self-regulation of the execution of these strategies can increase success in many domains, including software engineering, and there have been some informal descriptions of how to solve specific classes of software engineering problems, this prior work does not provide guidance on how to \emph{represent} programming strategies or guide developers on the use of them. This leaves an important gap in supporting and training software engineers.
\section{Explicit Programming Strategies}
\label{sec:ExplicitStrategies}

We define an explicit programming strategy as \textit{a human-executable procedure for accomplishing a programming task}. Figure \ref{figure:Barriers} shows one example of an explicit program strategy that provides step by step guidance for resolving a merge conflict. This example is \textit{human-executable} in that a human can interpret and follow its steps, and it is a \textit{procedure} in that the steps are a well-defined series of actions to be performed.

As with strategies in other domains, programming strategies are expressed imperatively, describing steps in which a human acts or retrieves and interprets information from the world. Describing a sequence of steps as a strategy does not mean it is the best approach or even that it is ever a successful approach; for example, there are other ways to resolve merge conflicts than the approach portrayed in Figure \ref{figure:Barriers}. For a given programming task, there may be many distinct strategies, which may vary in effectiveness; and, for each strategy, there may be many variants, which might choose to include or exclude details or handle specific cases in different ways. 

Explicit programming strategies capture detailed prescriptive advice describing \textit{how} to work in a specific way, going beyond high-level characterizations of the types of activities developers should do. For example, consider a high-level characterization of a software development process, such as agile software development. Agile specifies that working software should be preferred over comprehensive documentation and responding to change should be preferred over following a plan \cite{beck2001manifesto}. In practice, there may be countless ways developers might act that is consistent with this advice. A project might have an agile practice of never documenting anything under any circumstances but instead collocate software developers. This might then require a strategy describing how a developer might decide who to ask for a specific issue, how much, if any, investigation to do before asking them. There might be a family of agile strategies which describe an agile approach to the problem of eliciting requirements, actions a developer should take whenever a requirement changes, or how to decide what decisions, if any, should be documented. In this way, strategies might be used to teach a software development practice, but offer much more specific, prescriptive advice describing concrete actions a developer may take in the midst of a software engineering activity and the order in which they should occur. 

Programming strategies may be represented in a variety of forms. One might  characterize test-driven development \cite{beck2003test} as a process wherein a developer writes tests first, or more specifically as writing a test for a behavior, verifying that it fails, implementing the behavior, and then testing the behavior. But even this more specific description leaves some of this activity unspecified: when, if ever, should a developer identify and address design issues? Should these issues be addressed after implementing each behavior, or all together after implementing many behaviors as part of a larger feature? How and from where should the developer identify the behaviors? As descriptions of strategies become more detailed to offer guidance on such questions, pure natural language descriptions might become unwieldy or unreasonably ambiguous. This suggests the need for a representation of strategies that separately lists actions for a developer to take, just as in a recipe or checklist. In situations where developers act conditionally or repeat actions (e.g., edit the code to address the failure, run the tests, repeat until all the tests pass), it may be helpful to introduce control structures or to even decompose strategies into separate sub-strategies, to help developers be more thorough and systematic in following the strategy. As strategies offer more specific guidance, become more detailed, and grow in complexity, this suggests the need for a more formal and structured notation, such as a programming language.



A fundamental goal of a notation for programming strategies is to enable a developer to change their behavior to be consistent with a strategy.
Prior work on lab protocols, a form of mixed-initiative program in data science, found that it is important to offer both guidance as well as sufficient flexibility to accommodate expertise not captured in a procedure \cite{abbott2015programs}. We found similar requirements in an early prototype in which we simply presented an explicit strategy to the user as text and asked them to follow it. This high degree of autonomy introduced a number of challenges for developers. As they moved back and forth between the strategy text and their development environment, some lost their place and did not know which statement to read next. As developers' cognitive load increased due to the overhead of understanding the code, they sometimes forgot that they were being asked to work using a different strategy. Instead, they reverted to working through their traditional strategies, abandoning the remaining steps described in the strategy they were asked to use. These pilots, along with prior work, suggest that carefully choosing what kind of autonomy developers have when executing strategies is key to jointly leveraging human cognition and computation.

In this paper, we explore how we might describe explicit programming strategies for individual developers. We build upon recent work envisioning distributed human computation \cite{quinn2011human,abbott2015programs}, in which humans and computers jointly compute, requiring mixed-initiative interfaces that coordinate human and computer actions \cite{horvitz1999principles}. We bring these high-level visions for human-computer coordination to the specific domain of programming strategies. Our premise is that when working independently, a developer is left to plan and monitor plan execution while also reasoning about code, retrieving information, and devising design solutions. Novices lack many of these skills, which lessens their ability to be systematic in their reasoning \cite{loksa2016programming}, and experts often engage in similar externalization strategies to regulate planning and information persistence during a task \cite{robillard2004effective}. Moreover, prospective memory (their memory for future plans) is faulty, especially as people age, making external memory aids critical to avoid failure \cite{einstein1990normal}. Therefore, our hypothesis is that \emph{by explicitly offloading these planning and information persistence functions to the computer, both novices and experts may perform better on tasks by allowing them to focus on information retrieval and decision making}.

How then, should this collaboration between developer and computer be supported? There are many possible ways. A ``high autonomy'' approach might simply be to encourage developers to write down their strategies for problems they encounter, check them regularly, and adjust their plans accordingly. Prior work shows that such informal task scaffolding can improve rank novice behavior \cite{loksa2016programming}; perhaps it can also improve the performance of more experienced developers, or perhaps the lack of structure would result in minimal impact. A ``low autonomy'' approach might be to precisely structure developer behavior, prescribing exact strategies for the developer to follow at the level of precision that might be required for a computer to execute (like a Standard Operating Procedure). This extreme form of explicit strategy might leave no room for the developer's own knowledge or expertise; it might also cause them to reject the strategy because it leaves them no autonomy.

In this paper, we present a point in the design space that is a compromise between these extremes, delegating roles to the computer and the developer as best suited. With this flexibility, it allows strategy authors to make an informed choice about how to divide the labor. We call this strategy description language \textit{Roboto}. In the rest of this section, we describe the Roboto strategy description language and then present a tool for executing Roboto strategies.






\subsection{Roboto}

\begin{figure*}[]

\includegraphics[trim=72 640 72 74, clip, width=1.0\columnwidth, keepaspectratio]{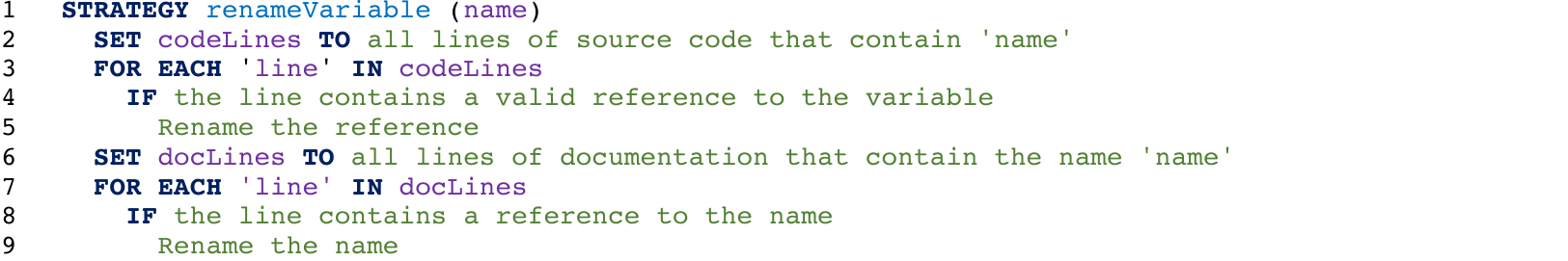} 


\caption{An example Roboto strategy that systematically renames a variable in source and documentation. In Roboto, the developer is responsible for retrieving complex information and making decisions. For example, the developer determines which lines of documentation reference the variable name. The computer helps ensure that the developer is thorough, systematic, and remembers past information they've written down. For example, the computer tracks the name of the variable to be renamed, the set of program statements a developer decides to inspect, and the next strategy step to execute, which is displayed to the developer through the strategy tracker. }	
\label{figure:example-strategy}
\end{figure*}

\begin{figure*}[]

\includegraphics[trim=72 640 72 74, clip, width=1.0\columnwidth, keepaspectratio]{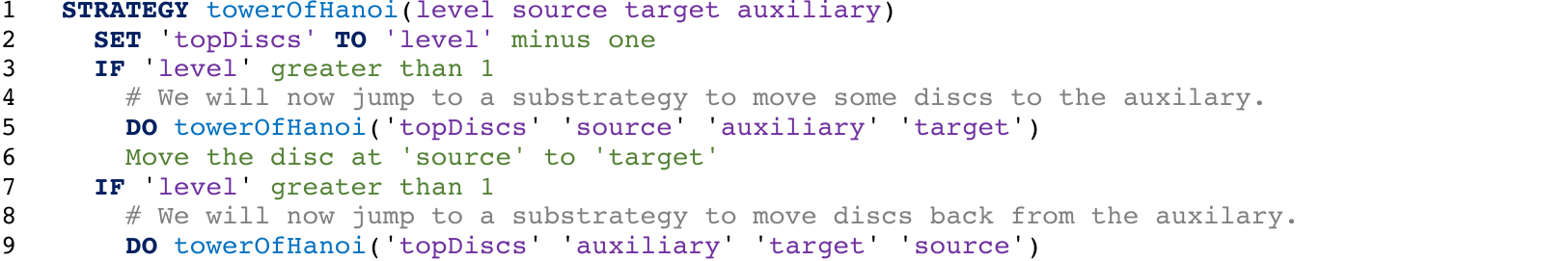} 

\caption{An example of a recursive Roboto strategy for solving the Tower of Hanoi puzzle.}	
\label{figure:example-strategy2}
\end{figure*}

The overarching design principle behind Roboto is that developers are better than computers at retrieving complex information and making decisions, while computers are better than developers at remembering, being thorough, and being systematic. For example, Figure \ref{figure:example-strategy} shows an example of a Roboto strategy that guides a developer through the error-prone process of renaming a variable in a web application built in a dynamic language. In this strategy, the computer is responsible for ensuring that the developer finds all of the lines to update, remembers to check documentation, and remembers to update each one; the developer is responsible for using tools to actually perform these actions in their editor or IDE. This notation gives developers the responsibility of acting in the world and retrieving information from it, and assigns the computer the responsibility for the strategy's control flow and its persistence of information from the world. Strategies may be short and simple or long and complex, may be organized into separate, cohesive substrategies, and may be recursive. While the focus of our design is on  strategies for programming, explicit strategies may also be used in other domains where computers and humans work together to accomplish a task. For example, Figure \ref{figure:example-strategy2} shows an example of a Roboto strategy for solving the Tower of Hanoi puzzle. Additional examples of explicit strategies can be found in our online repository \footnote{http://programmingstrategies.org}.

\begin{figure*}[]

\begin{verbatim}
STRATEGY :: strategy IDENTIFIER (IDENTIFIER+) STATEMENTS
STATEMENTS :: STATEMENT+
STATEMENT :: * (ACTION | CALL | CONDITIONAL | FOREACH | ASSIGNMENT | RETURN )+
ACTION :: (word | IDENTIFIER)+ .
CALL :: do identifier ( IDENTIFIER* )
CONDITIONAL :: if QUERY STATEMENTS
FOREACH :: for each IDENTIFIER identifier STATEMENTS
UNTIL :: until QUERY STATEMENTS
ASSIGNMENT :: set IDENTIFIER to QUERY
RETURN :: return QUERY
QUERY :: (word | IDENTIFIER | CALL)+
IDENTIFIER :: ' identifier '
\end{verbatim}

\caption{The Roboto language grammar.}	
\label{figure:grammar}
\end{figure*}

Figure \ref{figure:grammar} shows the Roboto grammar that achieves this division of responsibilities. Some of the non-terminals in the grammar convey actions that are executed by a developer:

\begin{itemize}
    \item \textbf{Actions} are a series of natural language words describing an action a developer should take in the world (e.g., \emph{``Rename the name``} in Figure \ref{figure:example-strategy}, line 9). Developers perform all of this work. Anything a developer would need to do, including operating a tool, navigating within an IDE, or asking a colleague for information, might be captured by an action.
    \item \textbf{Queries} describe information that a developer should obtain from the world, either to make a decision or to persist state for later use (e.g., \emph{``all lines of documentation that contain the name `name'``} in Figure \ref{figure:example-strategy}, line 2). This might include finding information in source code, conducting an Internet search, or some other activity producing information.
    \item \textbf{Comments} annotate a statement in a Roboto strategy and typically contain rationale for the statement or other knowledge a developer might need to successfully execute it (e.g., \emph{``We will now jump to a substrategy to move some discs to the auxilary``} in Figure \ref{figure:example-strategy2}, line 4). Comments are critical to convincing a developer that the next step is reasonable as well as helping developers understand the overarching approach of a strategy and how an individual statement supports that approach.
\end{itemize}    

The remaining non-terminals in the grammar are executed by the computer:

\begin{itemize}
    \item \textbf{Assignments} take the result of a query that a developer has executed and store the resulting information in a variable (e.g., \emph{``SET codeLines TO all lines of source code that contain name `name'``} in Figure \ref{figure:example-strategy}, line 2). Multiple assignments may reference the same variable, storing a new value which overwrites the previous value. This is effectively a way of persisting some fact for later use by the developer or the computer, much like a developer might use a notepad to keep notes during a task. A computer is responsible for this since it will not forget or lose the information a developer has captured. 
    \item \textbf{Conditions} take the result of a query and branch the strategy's execution (e.g., \emph{``IF the line contains a reference to the name``} in Figure \ref{figure:example-strategy}, line 8). During execution, they ask a developer to indicate whether the associated query is \texttt{true} or \texttt{false}, and then the computer branches accordingly. We gave this responsibility to the computer because of its larger responsibility in structuring the control flow of a strategy.
    \item \textbf{Loops} (\texttt{for each} and \texttt{until}) also involve a condition, and iterate through lists of data a developer has retrieved from the world or ask a developer to repetitively perform a series of actions until a condition is reached (e.g., \emph{``FOR EACH `line' in `codeLines'``} in Figure \ref{figure:example-strategy}, line 3). After finishing the execution of the statements within the loop's block, the computer returns control to the first statement. In this way, the computer is responsible for ensuring that developers are exhaustive in their consideration of data, rather than satisficing \cite{simon1972theories}, helping to prevent inefficiencies and defects in decision making.
    \item \textbf{Strategy} enables strategies to be functionally decomposed into sub-strategies and invoked, supporting functional decomposition and reuse of strategies, while also giving further control flow to the strategy (e.g., \emph{``STRATEGY renameVariable (name)``} in Figure \ref{figure:example-strategy}, line 1). Like in other programming languages, sub-strategies execute, then return control of execution to the caller. Sub-strategies also create separate variable scopes, reducing the information a developer must consider at once.
\end{itemize}

In allocating responsibility between the human and the computer, Roboto represents an intermediate point between pure natural language descriptions and traditional programming languages. Existing catalogs of strategies use natural language, describing strategies as a list of actions, relying on sentences of text to describe conditional behavior or repetition, or else omit conditional behavior and iteration entirely \cite{felleisen2001design,Kerievsky2004Refactoring,Fowler1999Refactoring}. In making explicit control flow and permitting strategies to be structured into sub-strategies, it is possible to more precisely describe exactly when particular actions will be taken and to group and reuse complex sequences of actions. This makes strategies more explicit by clarifying exactly what a developer should do as well as enabling better tool support for strategy execution. 

Roboto, as a notation, differs from programming languages in several ways by what it does \textit{not} include. Work that is allocated entirely to the human is left as natural language descriptions of actions. For example, there are no comparison operators, as the human makes judgments about the state of the world. Similarly, there are no arithmetic operators, as the human is responsible for acting in the world. And loops do not include increment forms, as either the computer has responsibility for ensuring all elements in a collection are visited in a for each loop or the human has responsibility for determining when a condition is reached in an until loop.

One complexity of persisting state through variables is whether to \textit{type} the results of queries that developers perform. In our design, all variables are either strings or lists of strings, under the assumption that most tasks will require the description of some state in the world, and that developers are best capable of writing these descriptions for their use. We support lists so that the computer can systematically iterate through items retrieved by a developer.

None of these design decisions are necessarily ``right.'' As we have discussed, there are many possible ways to describe explicit programming strategies, and Roboto is just one. Because our investigation was formative, our goal was not to find the ``best'' strategy notation, but to investigate the impact of ``a'' notation on developer behavior, building an understanding of how alternative decisions might have led to different behavior. We would expect many future works to explore these design choices in more detail.
\subsection{Supporting Strategy Execution}

\begin{figure*}[ht]
\vspace{0cm}
\includegraphics[width=\linewidth, keepaspectratio, clip]{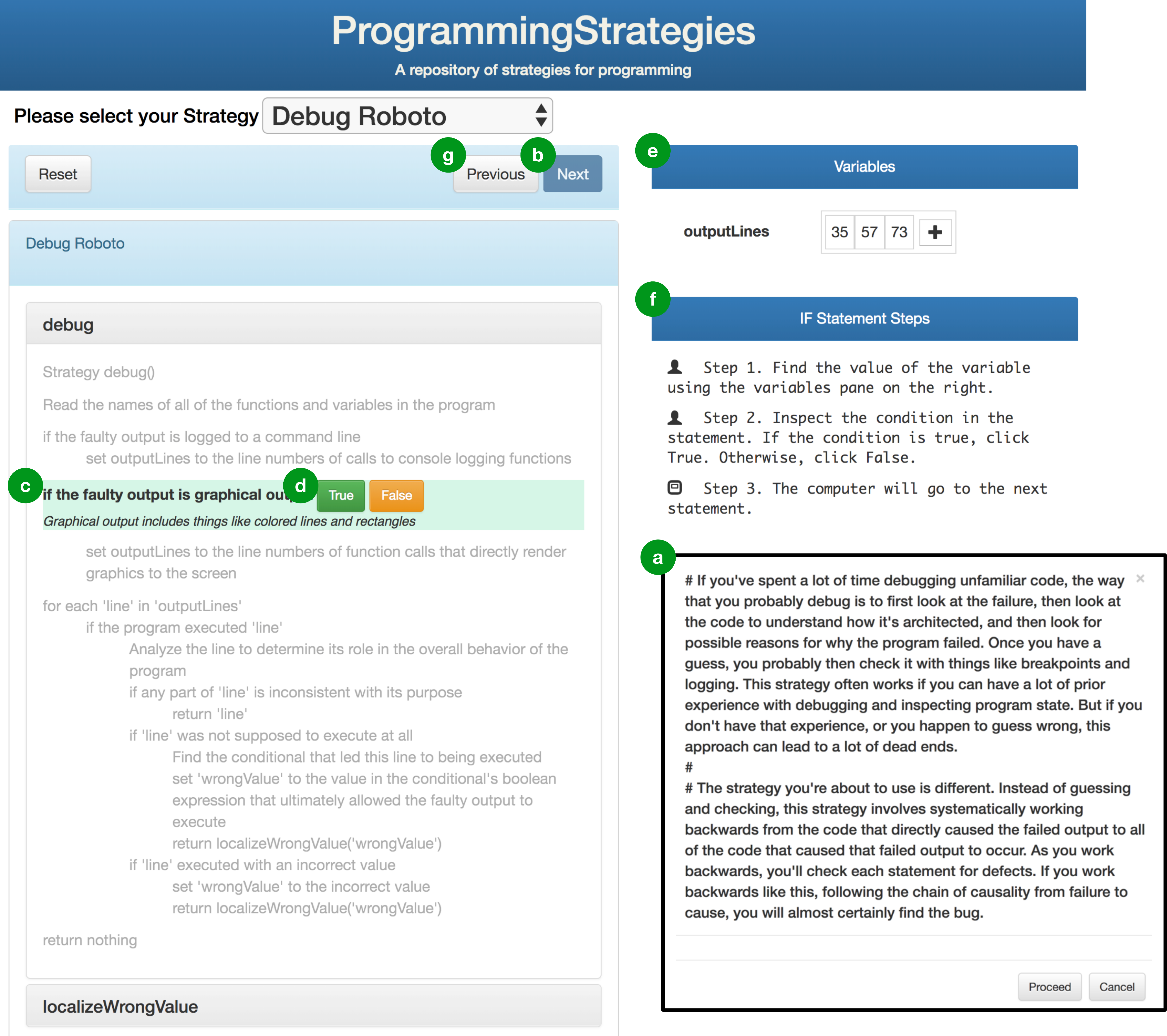} 
\caption{In the strategy tracker, the developer and computer work collaboratively to execute a strategy. After selecting a strategy, the developer (a) is introduced to the strategy through an introductory popup. The developer then works through the strategy statement by statement, (b) using the next button to advance the program counter and (c) performing the described statement, and (d) communicating decisions to the tracker when necessary. Developers may (e) record a value for a variable, offloading to the computer the responsibility of remembering it. For each statement, (f) the division of responsibility between the computer and developer is outlined through a list of steps each should perform.}	
\label{figure:ToolScreenshot}
\end{figure*}

Successfully following a strategy requires developers to be systematic by following each statement as written in a strategy step by step. We found in our early explorations that developers struggled to track the information they'd gathered from the world and keep their place in a strategy, which led them to not follow the strategy systematically. Therefore, to offload the responsibility of keeping track of a place within the strategy, we chose to design a strategy execution tool, adopting the model of an interactive debugger. The tool helps developers step through the strategy one statement at a time, executing the parts for which they are responsible, and letting the computer execute the parts for which it is responsible. Figure \ref{figure:ToolScreenshot} depicts the strategy tracker tool for Roboto.

When developers first begin executing a strategy, they need both confidence that the strategy they have selected is appropriate for their context and to understand what it will offer them. Roboto strategies may have an introductory comment, describing the purpose of the strategy. As a developer begins the strategy, this information is displayed to the developer in the form of a popup (Figure  \ref{figure:ToolScreenshot}a). When an initial strategy includes parameters, the popup displays the name of each parameter and prompts the developer to enter an argument value for each in a textbox (e.g., ''requirements'', with a textbox to enter a list of requirements).

Like in a traditional debugger, a program counter keeps track of the  active statement in the strategy. As in Roboto each statement constitutes an individual and indivisible operation, control passes from statement to statement. The developer advances the program counter with the next button (Figure  \ref{figure:ToolScreenshot}b), which steps the user to the next statement. The current statement is indicated to the developer by highlighting and bolding the statement text. This reduces the burden on the developer's working memory, enabling the developer returning their focus from the programming environment to the strategy to rapidly see where to resume.

To help developers understand the rationale for a statement, the tracker provides both the statement and a comment (listed in a line immediately below the statement), which may offer rationale and additional guidance (Figure  \ref{figure:ToolScreenshot}c). Developers who are more expert may focus only on the statement while ignoring the guidance, while the comment provides a learning resource for developers with less experience. 

For some Roboto statements, input from the developer is required before proceeding to the next statement. For \texttt{assignment} statements, for example, the tracker prompts the developer to enter a value for a variable, disabling the next button until they do so. For \texttt{condition} statements, the developer advances to the next statement by determining whether the query in the statement is \texttt{True} or \texttt{False} (Figure \ref{figure:ToolScreenshot}d), with the next button disabled.

Developers can assign values to variables using the variables pane (Figure  \ref{figure:ToolScreenshot}e), externalizing the value and offloading the burden of remembering it. When a \texttt{query} statement asks the developer to record a value, they can use the variables pane to write it down. All variable values are treated as a string or list of strings, giving the users flexibility to record data in whatever way is appropriate for the diversity of information they might gather from the world. Developers may also enter lists of values, separated with a comma, to denote multiple elements. When a list is referenced within a \texttt{for each} loop, the tracker automatically steps through each element, automatically assigning a named temporary variable a value corresponding to the current element in the list. To reduce information overload, variables are only shown in the variables pane after they have been referenced by at least one statement, as with a breakpoint debugger.

To reinforce the partnership between the computer and the developer, the developer needs a clear understanding of their role. Each statement in a strategy consists of steps executed by the computer and by the developer. A list of steps (Figure \ref{figure:ToolScreenshot}f) communicates, for the current statement, the responsibilities assigned to the computer and those assigned to the developer. Icons indicate whether each step is performed by the developer or the computer. For example, in a \texttt{conditional}, the developer is assigned the responsibility to find the value of any referenced variables, interpret the query to determine if it is true or false, and communicate this by clicking \texttt{True} or \texttt{False}. The computer then determines the next statement to execute and advances the program counter.

As in a breakpoint debugger, the strategy tracker maintains a stack frame for each sub-strategy, containing the values for local variables defined in the sub-strategy. When a developer enters a sub-strategy, a new stack frame is created for that sub-strategy. Variables that were in scope for the previously executing sub-strategy are hidden from the variables pane. When control returns to the calling sub-strategy, the stack frame is popped and the variables pane again shows the values for the calling sub-strategy.

To grant developers flexibility in how they use the strategy, the developer can, at any time, change the value of variables, including those not assigned in the current statement. If a developer later decides that the value they had written down in a previous step was incorrect, they can edit it. Changing the value of a variable only influences the execution of subsequent statements. If the developer intends to change the path taken, the developer may use the previous button to go backwards to before the conditional, edit the value, and then proceed. Loops pose an additional complication, as the developer might delete the currently active element in the loop or delete an element which has already been used in the loop iteration. For this reason, editing or deleting values previously used in the loop iteration is disabled. Developers may influence the subsequent behavior of the loop by editing, adding, or deleting values that have not yet been referenced.

To give further flexibility, the tracker also gives developers the ability to step backwards, as in a reversible or omniscient debugger \cite{pothier2009back,ko2004designing}. If a developer realizes that they have gone off track, making the wrong choice at a conditional or loop, they can use the previous button (Figure \ref{figure:ToolScreenshot}g) to move step by step back to the point where they diverged and then resume forward progress. When a developer steps backwards, they undo the statement they just executed and return to the state of the previous statement. If, in executing this statement, the developer had assigned a variable, either as instructed by an \texttt{assignment} statement or in editing any other variable's value, this action is undone and the value is reverted. Variables that have no longer been referenced in the current sub-strategy are again hidden. In this way, stepping backwards lets developers return to the state they were in before and try an alternative path forward.

As developers may complete programming tasks using diverse tools, including a wide variety of integrated development environments, programming languages, command line tools, communication tools, and websites, we designed the strategy tracker to be used alongside any programming tool.
To achieve this, we implemented the strategy tracker as a web application. 
Developers may interact with the strategy tracker in one window side by side with the programming tools they use to accomplish their task. Strategies may be referenced in a developer's work by, for example, including a link to a strategy within another tool (e.g., in a slack message or code commit), which may then open a browser window with the strategy. Programming tools may offer deeper integration by hosting the strategy tracker from a dedicated browser window within the tool, which might enable automatically configuring window positions to show the strategy and appropriate windows side by side.

\section{Formative Evaluation}
\label{sec:Evaluation}

Most of the empirical studies in software engineering research are \textit{summative} in nature, attempting to answer a question with some certainty by gathering and analyzing data. Summative studies tend to pose well-defined hypotheses, ideally derived from well-defined theories, and then test them. For example, a notable study of the effect of physical distance between developers on defect density in their components was a summative study, testing the theoretically-grounded hypothesis that distance matters \cite{bird2009does}. In other disciplines, such as medicine, summative studies are things like large human-subject clinical trials that carefully analyze the causal effects of some medical intervention such as a drug or surgery at scale.

In contrast, \textit{formative} studies, rather than seeking to rigorously \textit{test} hypotheses, seek to rigorously \textit{generate} hypotheses. Formative studies are typically done at the beginning of a field's investigation into a new phenomenon, to help identify what \textit{might} be true about a phenomenon, so that theories might be developed and hypotheses derived from that theory might be tested in future work. In software engineering, formative evaluations include things like case studies of new tool innovations, which do not demonstrate utility, value, or feasibility at scale, but rather identify existence proofs of value, while also surfacing unsolved problems. In medicine, a formative study might be an early, small-sample, animal-subject observational study, which reveals possible effects of an intervention for later study. Formative evaluations such as these are common in education research \cite{frick1999formative} and in HCI research \cite{sears2007human}, two disciplines our work builds upon.

Our evaluation of Roboto and the strategy tracker was \textit{formative} in nature: our goal was not to test whether Roboto and its tracker are ``effective,'' but rather to understand \textit{how} explicit strategies written in Roboto and executed in the tracker \textit{change} developers' strategic behavior and thereby help or hinder their progress, so that future work may improve upon the effectiveness, utility, and use of explicit programming strategies. Specifically, we sought to investigate:

\begin{itemize}
    \item What strategies do guided and self-guided developers choose to use?
    \item How do explicit strategies help and hinder developers' problem solving?
    \item To what extent do explicit strategies improve success on debugging and design tasks?
\end{itemize}

As Roboto and our tool for executing Roboto strategies are just one first attempt to explore explicit strategies, these questions help us identify hypotheses and questions for future work to investigate, not to provide definitive summative evidence of benefits. Because of this, we focused on closely analyzing a smaller set of developers' strategic behavior rather than a more shallow analysis of a larger group of developers' behavior.

The design of our study involved two groups completing the same set of two tasks. Our goal was for each group to have similar basic knowledge of a set of languages, APIs, and tools, but varying the independent variable of strategic knowledge. One group, which we will call \textit{self-guided}, completed the tasks through standard practice, unaided by explicit strategies or the strategy tracker tool. This group provided us a baseline for comparison, helping us understand variation of strategies without the presence of explicit Roboto strategies. The second group, which we will call \textit{guided}, completed the tasks with explicit strategies through the strategy tracker tool and was instructed to follow the strategies as long as they were helping accomplish the task. These two groups allowed us to contrast guided and self-guided strategic behavior, highlighting the strengths and weakness of both forms of programming work. Our study materials and replication package are publicly available \footnote{https://github.com/devuxd/ExplicitProgrammingStrategiesStudyMaterials}.

\subsection{Programming Strategies}

Rather than construct our own explicit strategies, we adapted strategies reported in prior work into Roboto notation, including those that have been fully automated in prior tools but are also appropriate for manual execution and strategies that cannot or have not yet been automated. We selected two strategies, one for design and one for fault localization.
 
\begin{figure*}[]

\includegraphics[trim=72 365 72 74, clip, width=1.0\columnwidth, keepaspectratio]{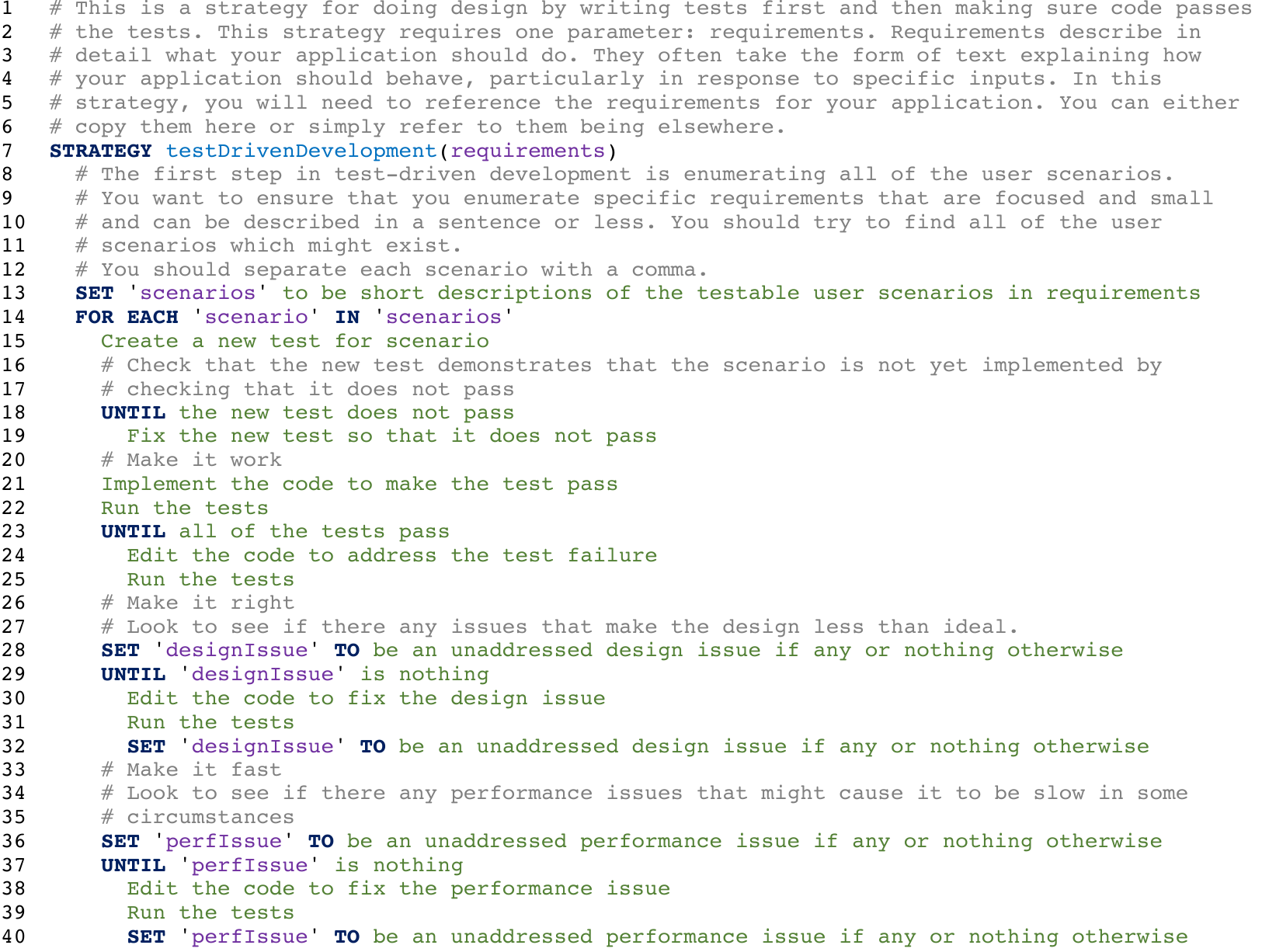} 

\caption{The test-driven development strategy, translating scenarios into tests, which drive development.}	
\label{figure:tdd-strategy}
\end{figure*}

For a design strategy, we adapted Test-Driven Development \cite{beck2003test} into the Roboto strategy shown in Figure \ref{figure:tdd-strategy}. We described each step in detail, adding comments to explain the motivation for each step. In piloting, we found that this explanation was largely sufficient in helping participants understand the idea of the strategy. 
 
\begin{figure*}[]

\includegraphics[trim=72 100 72 74, clip, width=1.0\columnwidth, keepaspectratio]{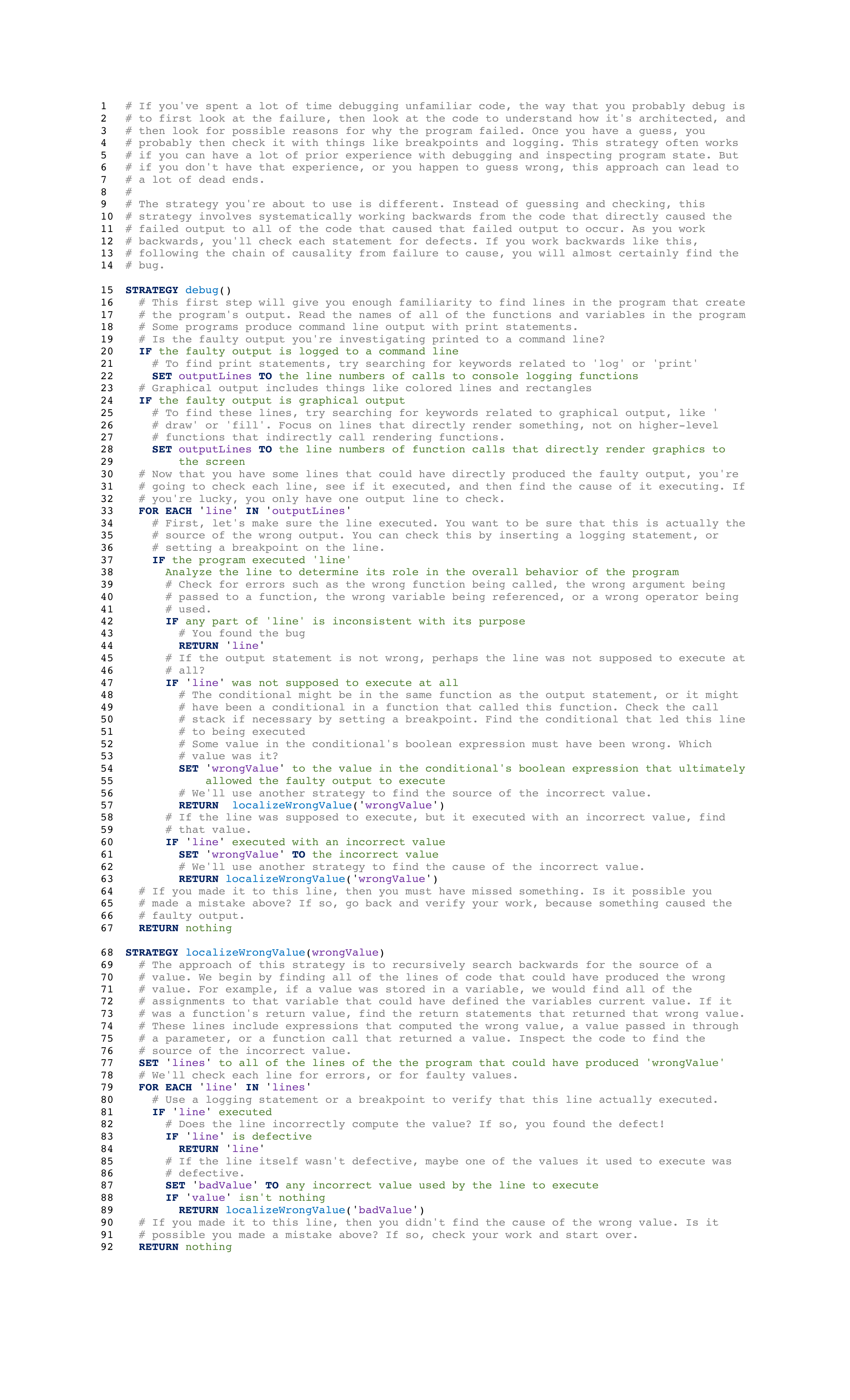} 

\caption{The backwards slicing debugging strategy, showing extensive comments detailing the rationale for each step.}	
\label{figure:debug-strategy}
\end{figure*}

For fault localization, we selected a precise backwards dynamic slicing algorithm, as used in the Whyline \cite{ko2010extracting}, which has been shown to significantly reduce developers' time to localize a fault \cite{ko2009finding}. While this strategy has been used in a purely automated tool, few widely used platforms support the execution tracing stack necessary to use the Whyline in practice. Therefore, manually following the algorithm by leveraging human cognition as a source of data collection and capture may be a reasonable substitute. Figure \ref{figure:debug-strategy} shows the Roboto strategy we wrote for precise backwards dynamic slicing. We found during piloting that in addition to providing the core approach to backwards slicing, the strategy also required extensive rationale and guidance to help developers execute their portion of the strategy successfully. We included these as comments, iteratively refining the explanations and rationale in the comments through pilot testing, while attempting to keep the strategy generic.

\subsection{Participants}

To understand the range of possible benefits and problems, we sought to recruit developers with diverse programming expertise. This would help us reveal the range of reactions that developers have to explicit strategies. Measuring programming expertise is still more of an art than a science, with no validated general instruments, and only early evidence of what programming expertise is and scant evidence of which factors correlate with it. However, recent prior work suggests that the best predictors of productivity and program quality are granular measurements related to a task, not years of experience \cite{dieste2017empirical}, and so we grounded our measures of expertise in the task domain we selected: front-end web development in JavaScript. Therefore, our inclusion criteria for study participation were robust knowledge of JavaScript semantics and robust knowledge of front-end web development APIs. For each, participants were given a JavaScript program and asked to describe its output. We scored each based on the number of correct described output lines, with a maximum possible score of 7 across the two tasks. There was a clear bimodal distribution in the middle of the scale, so we invited participants who scored 5 or above to participate. To ensure that participants in each condition had similar levels of expertise, we used stratified random sampling to assign participants to groups. For the 19 participants with the maximum possible score of 7, 10 were assigned to the control condition and 9 to the experimental condition. Of the remainder, 5 were assigned to the experimental condition and 4 the control condition.

    

Because our goal was a diverse sample with varying expertise, our recruitment strategy involved several distinct efforts. First, we recruited from undergraduate populations with students who had taken a web development course that taught HTML, CSS, JavaScript, and the React framework, as well as from undergraduate populations that self-reported experience in web development. Second, we recruited from several populations of graduate students, spanning both full time students as well as part time students currently employed full time as a software developer. Finally, we recruited full-time software developers from our professional networks. Overall, this yielded a sample of 28 participants, with a range of 2 to 28 years of programming experience (median 5, inter-quartile 5) and 0 to 15 years of industrial experience as a software developer (median 2, inter-quartile 3). Participants were between 20 and 36 years of age (median 25, inter-quartile 7). 21 participants were male, and 7 female.

\subsection{Tasks}

To help us explore the varying impact of explicit strategies, we asked participants to complete two tasks, one debugging task localizing a defect from reproduction steps and one design task translating a natural language problem statement into an implementation. Because our participants had a range of strategic expertise, we needed tasks that would be challenging for experts but not impossible for novices. Therefore, the debugging and design tasks focused on helping participants learn to use the strategies and strategy tracker, approximating first-time use rather than long-term expert use. 
To select these tasks, we first iterated on their design using pilot studies. The final debugging task provided a defective web-based action game with a snake that only moved diagonally rather than in response to keyboard input. 
The design task asked participants to create a simple auto-complete control which generates and ranks completions based on the words that the user has entered in a text area.


\subsection{Data} \label{sec:datacollection}

To understand how explicit programming strategies helped and hindered developers, we gathered several sources of data for analysis. 

Before beginning the study, we administered the inclusion criteria measure of JavaScript prior knowledge, then collected demographic data to characterize who was participating in the study. 

To explore the effect of prior expertise on progress, our approach was to focus on \textit{task} expertise, rather than overall programming expertise. This is because decades of prior work expertise in learning sciences clearly demonstrates that expertise is task-specific \cite{bransford2000people}. Therefore, to measure task-specific expertise, in our demographic survey, we presented natural language descriptions of the TDD and precise backwards dynamic slicing debugging strategies, and asked participants to self-report prior experience with each guided strategy (e.g., our TDD scale ranged from, ``used TDD extensively'' to ``have never heard of TDD''). While there are clearly many facets to expertise on these specific strategies, and likely many degrees of expertise with these facets, participants responses to these task-expertise questions were largely bimodal: most participants reported being entirely unfamiliar with each strategy, and participants who reported having using the strategy in the past. Therefore, we ultimately classified each participant, for each task, as either \textit{unfamiliar} or \textit{familiar} with the strategy. (We use these labels in presenting our results).

To observe participants' work, the experimenter had a display mirroring the participant's screen and described high-level visible actions the participant was taking, including interactions with the IDE, the strategy tracker, and transcripts of think aloud speech. While we initially asked participants to think aloud, participants often forgot, and we did not prompt them to think aloud more than once. We also collected the final code produced for each task as well as the time on task.

After completing each task, we interviewed participants, asking them to describe the strategies they used (\textit{``Describe the strategy or strategies you used to complete the task.''}) and how they helped and hindered their progress (\textit{``How did this strategy help in making progress on the task?''} and \textit{``In what ways, if any, did the strategy get in the way of making progress?''}). (We describe how we analyzed this data in Sections \ref{strategiesUsed} and \ref{strategyAttitudes}, where we present results).

While observing the actual actions that developers performed to execute their strategies would also be indicative of their strategies, and potentially more objective, prior work on problem solving in other domains (primarily mathematics) has shown that it is not possible to objectively infer strategies from actions because there are many possible different strategies that can lead to similar actions \cite{schoenfeld1981episodes}. A more common approach to observing strategy is to retrospectively ask a person to describe the strategies they used in as much detail as possible. While these retrospective accounts have the risk of not reflecting the strategies that someone actually used, prior work with children as young as 8 are capable of accurately describing their strategic intents \cite{desoete2001metacognition} as long as the prompt occurs immediately after a short period of problem solving. This has the added benefit of not requiring the experimenter to ask clarifying questions about the actions they were taking, which can invoke self-regulation skills that would not have otherwise been used \cite{chi1997quantifying}.

\subsection{Piloting}

A standard best practice in designing controlled experiments of software engineering tools is \textit{piloting}, which involves conducting a study procedure to identify and then eliminate confounding factors in measurement \cite{ko2015practical}. Because our study design involved a new idea of explicit strategies and a potentially unfamiliar programming environment, we piloted our study design and materials for the guided condition over 15 times before we had confidence that participants could comprehend the strategies and tools they were to use. Pilot participants were drawn from the same underlying population as the study participants, graduate and undergraduate students as well as full-time software developers with knowledge of web development. Each pilot participant completed both the debugging and design tasks. To identify points of confusion participants had understanding our materials, we used a number of signals: misuse of tools, misunderstanding of a strategy, misunderstanding of a task, usability problems with the tracker, misunderstanding of interview questions, and insufficient prior knowledge about the languages and platforms. After each round of piloting, we updated our materials to address these issues. After fifteen pilots, we no longer observed any of these categories of issues and were therefore confident that most participants would have sufficient time and cognitive resources to make sufficient progress on the tasks. We did not want or need \textit{everyone} to succeed at the tasks. This would have resulted in a ceiling effect in our task performance measurements, which would have masked the effects of the strategies we provided. We separately conducted three pilot sessions with the self-guided condition, who interacted with a subset of the materials used by guided participants.

\subsection{Procedure}

Prior to participating, participants completed the inclusion criteria assessment and those that met the criteria were invited to participate. Study sessions were conducted with individual participants both in-person (17) and remotely through screensharing (11).
At the beginning of the study session, participants completed a short demographic assessment. Participants then worked through a series of short tutorials on the WebStorm IDE, debugging in Chrome, and writing unit tests in the Jasmine testing framework and completed short exercises to apply their knowledge. Participants in the experimental condition were then introduced to the idea of programming strategies, the Roboto language, and the strategy tracker tool. The experimenter demoed using a strategy for handling merge conflicts in Git, demonstrating following the strategy using the strategy tracker. Participants then tried using a strategy themselves, solving a Tower of Hanoi example using the tool and the strategy in Figure \ref{figure:example-strategy2}. To minimize the disruption of switching between the tool and the Tower of Hanoi, participants were instructed to arrange each browser window side by side. Finally, participants were reminded that, to benefit from the strategy, they need to practice self-regulation and follow the strategy as described. 

All participants were then given the debugging task followed by the design task. Participants were given 30 minutes to complete each task. Participants in the self-guided condition were asked to solve the task, and participants in the guided condition were asked to use the strategy tracker to learn the strategy and apply it to the task. Guided participants were instructed to place the tool side by side with the development environment. At the end of the task, participants were interviewed about their strategy and their experience using the strategy. Participants were then given the design task and interviewed again about their experiences. Finally, participants in the experimental condition were debriefed about their experiences with explicit strategies and the strategy tracker tool. Participants were compensated with a \$30 gift card. The study was approved by our institutions' Institutional Review Boards.

\subsection{Results}

Our formative evaluation sought to answer three questions, which we answer through the data we gathered.

\subsubsection{What strategies did guided and self-guided developers use?} \label{strategiesUsed}

One of the first and most critical aspects of developer work was what strategies they used. Self-guided participants could choose any strategy, while guided participants had to decide whether to follow our prompt to use the provided Roboto strategies, or deviate from them. We expected to observe a diversity of strategies in both conditions.

There are no well-studied methods for identifying strategy from developer actions or verbal data. As noted in Section \ref{sec:datacollection}, we used a method from research on mathematics problem solving for eliciting strategies \cite{desoete2001metacognition}, prompting developers to \textit{retrospectively} describe in words the strategies that they used to solve the problem, and, if they used the provided strategy, to summarize it or refer to it directly.

To categorize these retrospective descriptions of strategy, we individually analyzed the strategy descriptions that developers provided for each task, using the transcripts of actions to help interpret and contextualize their descriptions. Our goal in this first analysis was to generate a set of distinct categories of strategies and a coding scheme for classifying them. We then used this coding scheme to categorize the one or more strategies that developers described using. To assess the reliability of the coding scheme, we had two of the authors independently categorize strategy descriptions for both tasks. Disagreements on the first pass were minimal and emerged from ambiguities in the coding scheme. After an additional pass, both authors reached 100\% agreement in which set of strategy codes to assign to each participant.

\begin{table*}
\caption{Strategies developers described using in the two tasks and their frequency by condition.}
\label{table:strategies}
\begin{tabularx}{\textwidth}{l|X|r|r}
\hline
Design & Description & Self-guided & Guided \\
\hline
Template & Found and used example code as a template for implementation. (\textit{``The first thing I did was to see the code and a template. One of things I want to do is to keep it simple. If I pick something too long or too complex, it make[s] it hard for me to modify it. 
'')} & 4/14 (29\%) & 0/14 (0\%) \\
Decompose & Analyzed functional requirements for sub-problems, implementing each independently \textit{``First I tried to determine what exactly the function is supposed to do and the input variables that are involved in it and what it is supposed to return just looking at the method signature... Then I... tried getting started on the function ... like creating any necessary objects and arrays...''} & 9/14 (64\%) & 0/14 (0\%) \\
TDD & Translated functional requirements into test cases, identifying sub-problems from test case requirements. \textit{``Once I'm implementing the tests for a scenario...
So you start by making a failing test which is essentially just creating a stub. And then you sort of fill out the stuff with the minimum amount of code that would just get it to pass.''} & 2/14 (14\%) & 11/14 (79\%) \\

\hline
Debugging & \\
\hline
Guess \& check & Participants found suspicious lines of code, modifying them and checking the effects of their modification. \textit{``At some point I did some experimentation, which is when I was looking for something different, I would change a line, I would add a line, I would copy a line and see what effects it would have. Just to help further my understanding of the code.''} & 2/14 (14\%) & 0/12 (0\%) \\
Forward search & Participants identified where the program began processing input, following its execution from there, analyzing it for defects. \textit{``I started by reading, getting an overview of the whole code in the IDE [to] see where the functions were. Then I looked for some functions that had to do with movement or input. Then [I] tr[ied] to understand two functions that I identified as problematic that would be causing the problem, which were the `check-input' and `move'.''} & 13/14 (93\%) & 3/14 (21\%) \\
Backward search & Participants identified faulty output and worked backwards through control and data flow dependencies to localize the statement(s) that caused the failure (same as the explicit debugging strategy provided to the guided group). \textit{``The approach I took was to... look at each line that put a graphic output on the screen. And then, run through each line really systematically to see like what it did, if it was working, if it was supposed to execute in the first place.''} & 1/14 (7\%) & 13/14 (100\%) \\

\hline
\end{tabularx}
\end{table*}

Table \ref{table:strategies} lists the categories of strategies that emerged from our qualitative analysis and examples of how developers described them. To our surprise, the strategies that developers used were \emph{not} particularly diverse. For the \emph{design} problem, the three strategies involved different approaches to decomposition: reusing the decomposition in an existing example program (\textit{template}), analyzing the problem for ideas for decomposition (\textit{decompose}), or using tests to drive decomposition (\textit{TDD}). For the \emph{debugging} task, participants either modified the program to understand its behavior (\textit{guess \& check}), searched forward from user input (\textit{forward}), or searched backward from faulty output (\textit{backward}). The forward and backward strategies are consistent with existing observations of debugging behavior \cite{Bohme2017Debug}.

Table \ref{table:strategies} also shows the frequency of use of each type of strategy by condition. Only 4 participants across both used more than one strategy while debugging. For the design task, most self-guided participants used the \textit{decomposition} strategy, with only a few following \textit{TDD} and \textit{template} strategies. Most of the participants in the guided condition reported following the \textit{TDD} strategy we provided. The one who did not had trouble describing a strategy (and for two others we were missing data because of corrupted audio files). None of the participants for which we had data used more than one strategy during the design task. 
For the debugging task, most self-guided participants used a \textit{forward} strategy, while two also used a \emph{guess \& check} strategy. All guided participants using the \textit{backwards} strategy we provided, except for one for which we had no data due to a corrupted audio file. A few interleaved a \textit{forward} search strategy into their process.

These results show that when developers retrospectively described their strategies for these two tasks, 1) there was great regularity in the strategies they reported using, 2) that developers given explicit strategies largely did not deviate from them, and 3) that, at least during 30-minute tasks, developers tend not to use multiple strategies.

\subsubsection{How did explicit strategies help and hinder developers' problem solving?} \label{strategyAttitudes}

Whereas the previous section described what developers \textit{did}, here we analyze developers' perceptions of how the strategies they used (theirs or the strategies we provided) influenced their problem solving process. To perform this analysis, we used the answers that developers provided to our post-task prompts of what ``helped'' and ``hindered'' their progress on their task. We gave the same prompt to both guided and self-guided participants, asking them to reflect on \textit{all} of the strategies they employed during the task. After transcribing their answers, we inductively developed a set of attitudes expressed across the answers in both
conditions. This resulted in a separate code book of positive and negative attitudes for each task, including 18 distinct attitudes about the debugging task, and 19 distinct attitudes about the design task.

Common practice in qualitative software engineering research is often to \textit{quantify} qualitative data by coding, counting it, and measuring inter-rater reliability in these counts. However, we took a different approach to qualitative coding recently advocated by Hammer and Berland, and now widely adopted in the learning sciences \cite{hammer2014confusing}. Their perspective on qualitative data argues that the richest meaning of qualitative data emerges from a detailed analysis of the \textit{disagreements} in a group of coders about what the data means and how it was classified according to a coding scheme. Therefore, rather than relying on inter-rater reliability measures for validity, we instead analyzed our disagreements to assess validity. To do this, our process for coding was to reach 100\% agreement on the attitudes in each response, and in the process, surface detail about our disagreements that revealed potential flaws in the validity of our coding scheme. To achieve this, two authors independently assigned zero or more help and hinder attitudes to the 108 transcribed responses (28 participants, two tasks each, two questions each). After this process, the two authors disagreed on only 12 of the responses. They discussed each of these 12 disagreements, some of which were attitudes that were overlooked and therefore added to the coding scheme, and some of which were disagreements about what counted as an ``attitude''. The coders discussed all other agreed upon codes and determined that their interpretations emerged from the same meaning of the data.

\textbf{Template strategies for the design task}. For the design task, the self-guided participants had similar attitudes about their chosen strategies. The 4 that used the \textit{Template} strategy (one of which reported no experience with TDD), for example, all described the examples they found as providing a clear starting point for their work, but that trying to comprehend the example code was challenging, and that this comprehension took extra time. For example, one participant said:

\begin{quote}
``At first, I thought it was kind of cheating. I don’t use library. At the beginning it makes you spend some time but at the end when you get that done and you will be able to add every case... For me trying to understand somebody else code is difficult... The way they named variables, the way they put the structure... it sometimes hindered progress.''
\end{quote}

Of the four developers using the template strategy, only one satisfied any requirements for the task.

\textbf{Decomposition strategies for design task}. Of the 9 participants (all self-guided) that reported using a \textit{Decomposition} strategy, 4 reported that it ``organized'' their work and another 4 reported that it gave them a ``starting point'' for their work. Of these 9, however, three reported that they viewed decomposition as taking extra time and one noted that mistakes in decomposition eventually required them to redesign their solution. One representative participant said:

\begin{quote}
``[Decomposition] was helpful because initially when I read the problem I thought I might be out of depth in this thing because I didn't understand how will I proceed with this problem, but after that, even I asked a few questions, I tried to understand the full specification of the problem so I can break it down into more understandable, more manageable problems... It helped me comprehend the problem a lot better and to approach the problem in a better manner.''
\end{quote}

All but one of these 9 participants using the decomposition strategy satisfied three or more of the task requirements.

\textbf{TDD strategies for the design task}. The 2 self-guided participants and the 11 guided participants that reported using a test-driven development strategy reported that TDD helped them ``organize their work'' (8 of 13), helped build ``confidence'' in the correctness of their code (3 of 13), helped avoid extra work (2 of 13), and provided a ``starting point'' (2 of 13). Attitudes between the guided and self-guided TDD participants about how TDD helped were indistinguishable, but sentiments about how it hindered differed: the two self-guided participants reported TDD taking extra time (2 of 2). Most guided participants reported no hindrances, except for some (2 of 11) reporting some uncertainty about how to interpret the explicit strategy's instructions and one reporting a fear of diverging from the strategy. One guided participant captured these attitudes well:

\begin{quote}
``[Test-driven development] was really good at, don’t overwhelm yourself with the details [you] don't need quite yet. By starting very simple, just getting it so there's... return an empty array of that, I make a little bit of progress. And for me, TDD is always a little bit like a game...
I always have a little incremental improvement that I could do on this task. That really helped.''
\end{quote}

While all participants using TDD felt like their work was more systematic, only those reporting experience with TDD made progress on implementing functional requirements. Those reporting no experience with TDD reported investing more of their time learning to plan and writing tests than writing code to satisfy tests.

\textbf{Guess \& check strategies for the debugging task.} The two participants who used the \textit{Guess \& check} strategy (both in the self-guided condition and neither successful at the task), mentioned that trying to debug by modifying the program caused them to focus too narrowly in their search. Both of these participants quickly abandoned this strategy for the forward search strategy.

\textbf{Forward strategies for the debugging task.} The 13 self-guided and 3 guided participants that used a \textit{Forward search} strategy reported that there was little helpful about it, with the exception of two participants that said it helped them gain some familiarity with the code. Many reported that it caused them to have a long startup period (4 of 16), that it did not feel systematic (3 of 16), and that it wasted time (2 of 16). These attitudes likely were influenced by most of these participants not localizing the defect. This developer captured these attitudes well:

\begin{quote}
``[I] felt scattered sometimes. Like I would go to one thing, and I feel like I get a little off track like 'what was I going here for?'... Or like once I eliminated that track, I had to like think of what I was doing before I went down that path.''
\end{quote}

Of these 16 participants who used a forward strategy, 7 reported before the task that they had experience with working backwards, but none used the strategy in the task. Moreover, the forward strategy was ultimately unhelpful, with only 3 finding the defect and none fixing it.

\textbf{Backward strategies for the debugging task.} The 14 participants that used the \textit{Backward search} strategy (all in the guided condition, but only 3 reporting prior experience with a backward search strategy) felt positive about the strategy and the explicit support for executing it. Many said it provided them a helpful sequence of steps to localize the fault (5 of 14), helped them be systematic in their process (2 of 14), provided them context for their fault localization (3 of 14), and saved time (2 of 14). All participants reported that it was helpful, including the 5 who made no progress on finding or fixing the defect. The only hindrances that the guided participants noted were that the explicit strategy occasionally had unclear instructions (2 of 14), and that they often felt there was not room to deviate using their instincts or experience (2 of 14). One participant described these trade-offs well:

\begin{quote}
``I don't typically do the due diligence of reading all of the variable names and function names when I'm dealing with this sort of thing. And it seemed pretty clear to me that this is maybe a really good idea. Because one thing I noticed was that my initial instinct was to try to really close[ly] read the flow of the program. Then, when I remembered that the task was actually just to read the variable names and function names, I was able to get through it much much faster. I still had a pretty good idea of actually how it worked without getting quite as in detail with the rest of the flow of the program... 
''
\end{quote}

This same guided participant shared their thoughts on how the explicit strategy hindered them:

\begin{quote}
``I'm following these instructions, and I'm trying very hard to adhere to the instructions. It can feel kind of confining... It made me kind of hesitant to trust my instinct on things... I felt the need to really slow down, read the instructions multiple times, and to not do any action that could sort of mess up my adherence to the instructions.''
\end{quote}

In contrast, the one self-guided participant that attempted to use a backwards search strategy said:

\begin{quote}
``Since I wasn't used to this code, at first it was a little overwhelming. I was like oh what does all this mean... I couldn't understand all of the key components interact with each other and how each of these functions or variables were being used...
\end{quote}

The overarching trend in these attitudes was that participants tended not to choose TDD and backward search when given a choice, even when they reported experience with them, but when compelled to use them, appreciated how explicit strategies helped organize their work and remember to take key steps in developing their understanding of a problem or program.




 

\subsubsection{To what extent did explicit strategies improve progress on debugging and design tasks?}

Whereas the prior two result sections show that most participants did not choose the TDD or backwards search strategies independently but found value when they used them, here we examine if explicit strategies helped developers. To assess outcomes, we used a separate definition of task progress for the design and debugging tasks.

\begin{figure}[H]
\begin{subfigure}{\textwidth}
  \centering
  \includegraphics[width=.7\linewidth]{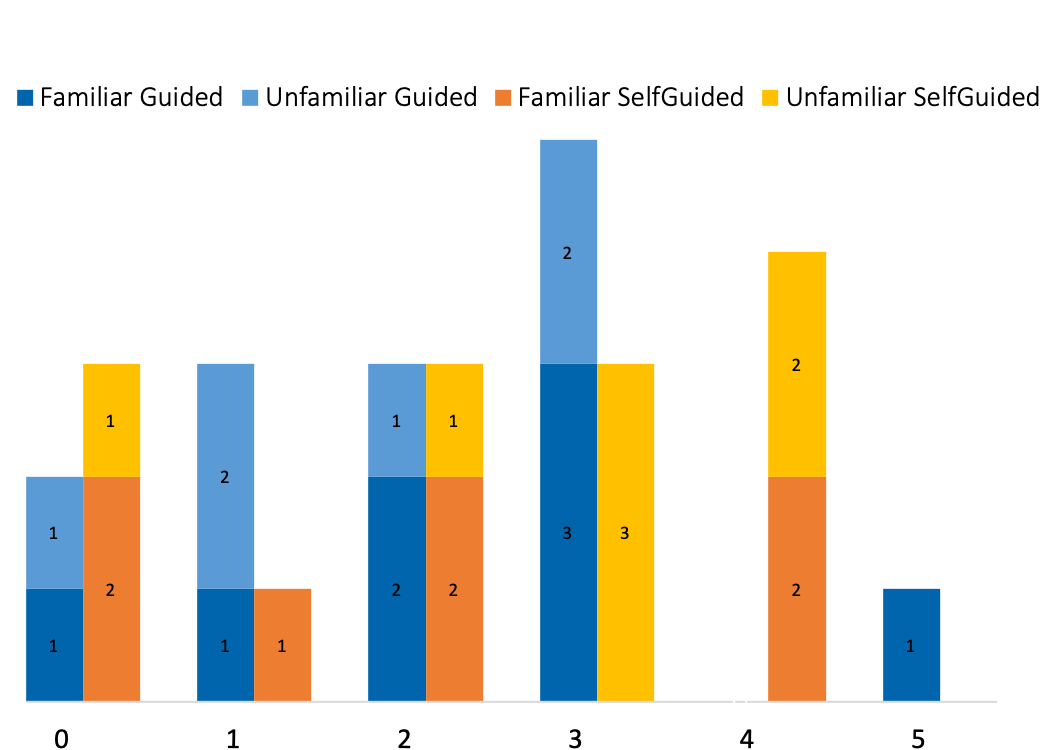}
  \caption{a) The number of requirements satisfied by participants in the design task }
  \label{fig:sub-reuirements}
\end{subfigure}
\begin{subfigure}{\textwidth}
  \centering
  \includegraphics[width=.7\linewidth]{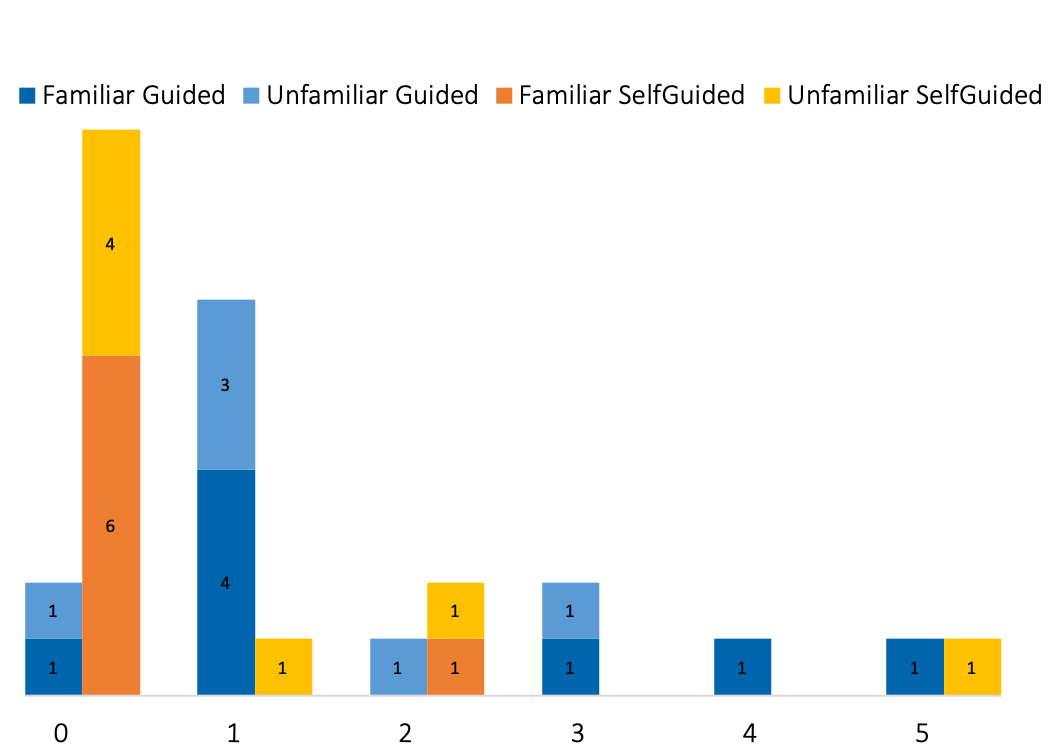}
  \caption{b) The number of relevant tests created by participants in the design task}
  \label{fig:sub-test}
\end{subfigure}%

\caption{Progress on the design task by condition and strategy familiarity. Each block counts the number of participants that attained each level of progress. }
\label{fig:designProgress}
\end{figure}

For the \emph{design} task,
we defined two independent factors for measuring the amount of progress. The first factor was how \emph{functionally correct} a participants' solution was with respect to our prompt. To measure this, we enumerated the requirements provided in the task description and developed a rubric for judging whether each of 5 requirements were satisfied or not. Two authors analyzed the problem statement for requirements, discussed ways of detecting having met these requirements in participants' implementations, and agreed upon a rubric that assigned a point per requirement, resulting in an ordinal 0-5 scale. One author scored each submission using this scale. Figure \ref{fig:sub-reuirements} shows the number of participant on each group of guided and self guided participants based on strategy familiarity and condition.

Our second measure of progress on the design task was the \emph{maturity of the solution's verification infrastructure}, which both guided and self-guided participants wrote. We counted each test that did not have syntax errors and that had a purpose related to a requirement, independent of a corresponding implementation. This resulted in an ordinal scale ranging from 0 to an observed maximum of 5 tests. Figure \ref{fig:sub-test} shows progress scores by condition and strategy familiarity for the number of total  participant on each group.

For the \emph{debugging} task, we defined three ordinal levels of progress: \textit{fixed} if they identified the location of the defect and proposed a correct fix, \textit{found} if they found the defect but could not fix it, and \textit{failed} otherwise. Our ordinal scale ordered these from low to high as \textit{failed}, \textit{found}, and \textit{fixed}.



\begin{figure}[H]
\centering
  \includegraphics[width=.7\linewidth]{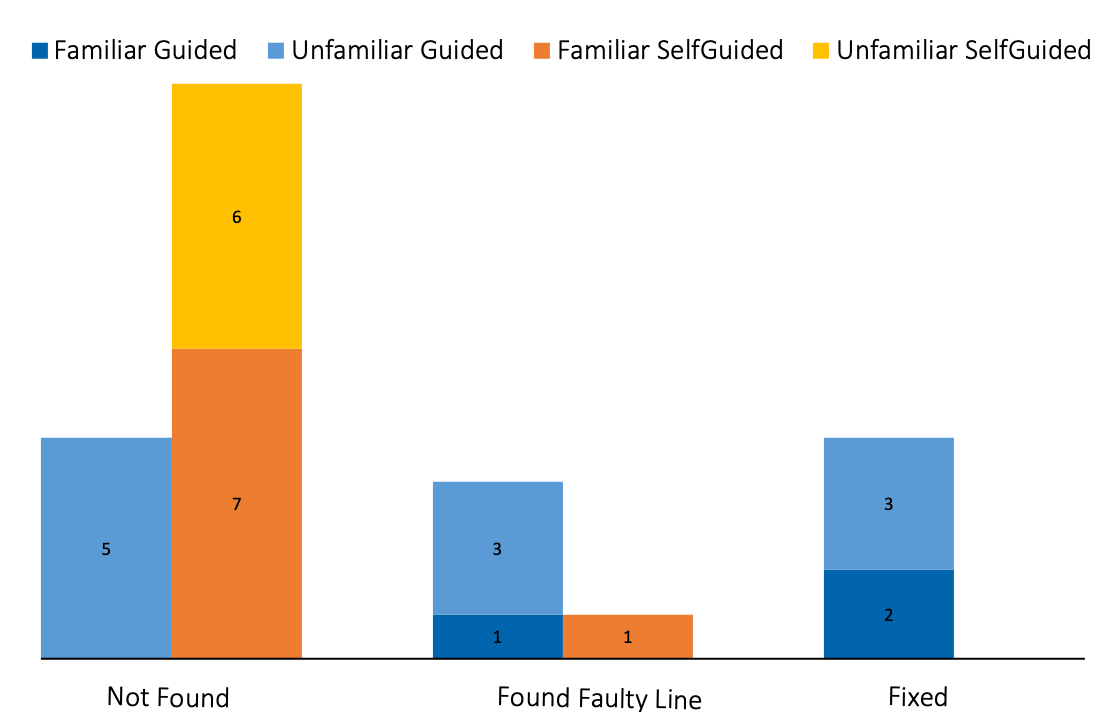}
  \caption{Progress on the debugging task by condition and strategy familiarity.}
  \label{fig:debuggingProgress}

\end{figure}

Figures \ref{fig:designProgress} and \ref{fig:debuggingProgress} show the level of progress that participants reached on each of the tasks. To examine if the explicit strategy manipulation or strategy familiarity had an effect on our three ordinal progress scales, we used the Wilcoxon Rank Sum Test, a non-parametric test to compare outcomes between two independent groups suitable for ordinal data \cite{hilton1996appropriateness}. Table \ref{table:wilcoxon} shows the results. Guided participants in the design task wrote significantly more tests, but did not satisfy significantly more requirements in their implementation. In the debugging task, guided participants made significantly more progress. Strategy familiarity with TDD was not related to progress on the design task with tests or requirements, and strategy familiarity with a backwards search debugging strategy was not related to progress on localizing the fault, suggesting that it was the explicit support for the strategy that resulted in more tests and more successful fault localization.

\begin{table}
\caption{Effects of guidance and strategy familiarity on task progress. Wilcoxon Rank Sum Test *=p\textless.05.}
\label{table:wilcoxon}
\begin{tabular}{l l l l }
\toprule
Task & Param & Diff & P-value \\
\hline
\multirow{2}{*}{Design-Implementation} 
&\multicolumn{1}{l}{Familiar} &  \multicolumn{1}{l}{87.0}& 
\multicolumn{1}{l}{0.3021}\\
& Guided & 
\multicolumn{1}{l}{82.5}& 
\multicolumn{1}{l}{0.2325}\\
\multirow{2}{*}{Design-Tests }
&\multicolumn{1}{l}{Familiar} & 
\multicolumn{1}{l}{72.0} & 
\multicolumn{1}{l}{0.1036}\\
& Guided & 
\multicolumn{1}{l}{48.0}& 
\multicolumn{1}{l}{0.0076*}\\
\hline
\multirow{2}{*}{Debug} &\multicolumn{1}{l}{Familiar} &  
\multicolumn{1}{l}{92.5}& 
\multicolumn{1}{l}{0.4779}\\
& Guided & 
\multicolumn{1}{l}{39.5} & 
\multicolumn{1}{l}{0.0008*}\\\hline
\end{tabular}
\end{table}

To understand the impact of the explicit strategy manipulation and strategy familiarity jointly, we built an ordinal logistic regression model, which controls for each factor. We chose an ordinal logistic regression model as the dependent variable, progress, in both tasks was ordinal. 
Table \ref{table:regression} lists the model parameters. The model shows that, accounting for the effects of strategy familiarity on TDD, guided participants in the design task were 1.3 times as likely to write more tests. In the debugging task, guided developers were 1.96 times as likely to make more progress on localizing the fault. Notably, while choice of strategy was significantly associated with more tests and successful debugging, strategy familiarity was \textit{not}.

We also analyzed the effects of explicit strategies on task time. Overall, 23 out of 28 participants on the debugging task used the full task time and 22 out of 28 participants on the design task used the full task time. Analyzing time by condition on the debugging task, 4 of the 5 participants who finished early were guided. On the design task, 5 of the 6 participants who finished early were guided. We tested for an effect of the explicit strategy manipulation and strategy familiarity on task time with a Wilcoxon Sum Rank Test. As the results in Table \ref{table:timeWilcoxon} show, the effects of strategy familiarity on task time was not significant for either task. The effect of guidance on task time was not significant for the design task and approached significance for the debugging task (p = 0.056).


\begin{table}
\caption{Effects of guidance and strategy familiarity on task time. Wilcoxon Rank Sum Test *=p\textless.05.}
\label{table:timeWilcoxon}
\begin{tabular}{l l l l }
\toprule
Task & Param & Diff & P-value \\
\hline
\multirow{2}{*}{Design} 
&\multicolumn{1}{l}{Familiar} &  \multicolumn{1}{l}{93.5}& 
\multicolumn{1}{l}{0.393}\\
& Guided & 
\multicolumn{1}{l}{75.5}& 
\multicolumn{1}{l}{0.870}\\

\hline
\multirow{2}{*}{Debug} &\multicolumn{1}{l}{Familiar} &  
\multicolumn{1}{l}{93.5}& 
\multicolumn{1}{l}{0.500}\\
& Guided & 
\multicolumn{1}{l}{75} & 
\multicolumn{1}{l}{0.056}\\\hline
\end{tabular}
\end{table}

\begin{table}
\caption{Effects of guidance and strategy familiarity on progress. Ordinal logistic regressions for each task. *=p\textless.05.}
\label{table:regression}
\begin{tabular}{l l l l l l l l}
\toprule
Task & Param & Odds ratio &  
SE B & Wald  & Pr $> {\chi}^2$ \\
\hline
\multirow{2}{*}{Design-Impl} 
&\multicolumn{1}{l}{Familiar} &
\multicolumn{1}{l}{0.60} &
\multicolumn{1}{l}{0.682}& 
\multicolumn{1}{l}{0.204}& 
\multicolumn{1}{l}{0.651} \\
 & Guided & 
\multicolumn{1}{l}{0.73} &
\multicolumn{1}{l}{0.691} &
\multicolumn{1}{l}{0.487}& 
\multicolumn{1}{l}{0.485}\\
\multirow{2}{*}{Design-Tests}
&\multicolumn{1}{l}{Familiar} &
\multicolumn{1}{l}{0.84} &
\multicolumn{1}{l}{0.752} & 
\multicolumn{1}{l}{0.808} & 
\multicolumn{1}{l}{0.369} \\
& Guided & 
\multicolumn{1}{l}{1.30 } &
\multicolumn{1}{l}{0.799 }& 
\multicolumn{1}{l}{5.177}& 
\multicolumn{1}{l}{0.023*}\\
\hline
\multirow{2}{*}{Debug} &
\multicolumn{1}{l}{Familiar}&
\multicolumn{1}{l}{1.40} & 
\multicolumn{1}{l}{1.296}& 
\multicolumn{1}{l}{2.807}& 
\multicolumn{1}{l}{0.094}\\
& Guided & 
\multicolumn{1}{l}{1.96} &
\multicolumn{1}{l}{1.562} & 
\multicolumn{1}{l}{8.398}& 
\multicolumn{1}{l}{0.004*}\\\hline
\end{tabular}
\end{table}


 
\section{Threats to validity}

As with any empirical study, our study design had several types of threats to validity.

There were several issues in our study design related to construct validity. Because there are no well-validated measures of prior knowledge in programming or specific strategies, our measures were coarse, which means that our claims about the relationship between task expertise and task performance are tentative. In measuring progress on the debugging and design tasks, we chose measures that assessed how developers localized the defect for the debugging task and how many requirements and tests of these requirements developers wrote for the design task. It is possible that developers may have made some progress in ways that were not measured, for example formulating a correct hypothesis about the cause of a defect without a location or formulating a plan in their head about how to implement a behavior without writing it down. To identify the strategy that each participant used, we grouped similar strategies into clusters. As with any clustering approach, strategies might have been further broken down into additional sub-clusters reflecting variants of the higher-level strategies we focused on.

From a conclusion validity perspective, there were several issues. First and foremost, our sample included diverse programming expertise, but this diversity inevitably led to significant variation in task performance. This decreased our ability to see the effects of explicit strategies experimentally. Additionally, calibrating task difficulty is always challenging, especially with such diverse prior knowledge. This meant that both the TDD and debugging tasks had floor effects, with many participants making no progress in the short amount of time we offered them. This reduced the sensitivity of our measurements, reducing our ability to precisely assess the effect size of our results. To ensure that incentives did not bias participants, all participants were compensated the same amount, regardless of condition, task performance, or their responses.

From an external validity perspective, our study design had many artificial qualities. Our participants were unfamiliar with the concept of a programming strategy, and we provided training materials as well as a demo by the experimenter. In practice, developers who were already familiar with the concept of a strategy might not need as extensive training materials. Conversely, training inexperienced developers might require more scalable training materials, without the need of in-person training.

In practice, we would expect developers to learn strategies over a much longer period of time and eventually excel at applying them (perhaps even without the aid of the strategy execution support we provided). In our experiment, however, we measured only the first exposure to explicit strategies, with very little time for practice. While this is an important part of strategy use to study---after all, any adopter of explicit programming strategies will go through it, and this determines whether they are likely to use them again---it does not address the potential of explicit strategies to shape programming practices over long periods of time in professional contexts. Our sample may also have been biased, since the authors' recruited participants from their social networks. There may be unique demographics in these groups that do not capture programming expertise in other social networks. The two tasks that we chose, while representative of the broad categories of design and debugging tasks, are specific tasks that do not represent the full range of task variation in software engineering. Investigating the effect of explicit strategies on that range is necessarily left to future work. Finally, we chose tasks for which developers could make progress within the time bounds of a lab study. We did not observe developers working on longer and more complex debugging or design tasks. These might raise additional challenges which might increase or decrease the benefits offered by explicit strategies.

\section{Discussion}
\label{sec:Discussion}
Our formative evaluation illustrates some of the potential benefits of explicit programming strategies. Developers guided by a strategy described their work as more organized, systematic, and predictable; they also performed objectively better on debugging and testing tasks. Guided developers worked qualitatively differently than the ways in which unguided developers naturally worked, debugging through backwards rather than  forwards search and designing using test-driven development rather than simple decomposition. In this way, developers were able to change the way they approached these problems and benefited from the collective software engineering wisdom embodied in these strategies, unlike the unguided developers, who had to rely on only their own strategic expertise. 

Our initial results raise a variety of important questions in how developers might effectively interact with a programming strategy. Future studies might include multiple conditions with different strategies to compare their effectiveness, employ a longitudinal deployment of a strategy into an organization to examine its use and adoption over time, or examine how developers might create strategies by asking experienced developers to write down their own strategies. 

The effectiveness of a developer employing a novel strategy depends on the context in which it is used. In our study, we found that developer who adopted a backwards strategy were more effective than those who used a forward strategy. But one might imagine a task for which the opposite is true, if, for example, the defect is located closer to the input processing than the output processing. Whether or not test-driven development results in overall productivity benefits may also depend on context, as its benefits are not consistently visible \cite{Shull2010TDD}. In this way, much more research is necessary to understand the fitness to purpose of various software engineering strategies for the typical situations in which developers work. Making strategies explicit and capturing  strategies is an important first step, helping open up a future space of research that examines how, when, and whether specific strategies are effective. If the number of strategies available were to increase, helping developers choose an effective strategy based on their context would also become a critical problem.

While we believe Roboto is an interesting point in the design space of programming strategy description languages, it is not the only possible design. For example, one trade off of giving power over all control flow to the strategy is that when a developer has better information than the strategy, or a better strategy altogether, a developer must effectively abandon the explicit strategy altogether and go forward alone. Strategy languages that make different choices from Roboto might better support reuse of persisted state and points of deviation in a strategy flow. Prior work on organizational behavior also suggests that any sufficiently complex goal requires open-ended adaption. Crowdsourcing researchers have argued that this is an inherent limitation of workflows built into crowdsourcing systems, suggesting that explicit strategies might be powerful for some forms of tasks but not necessarily all tasks, and especially not for tasks involving entire teams and organizations. \cite{retelny2017no}. It might be possible to create alternative language designs that are more reactive, where the developer has more control to change the plan they are following through mechanisms such as events or exceptions. However, our study participants found value in being encouraged to be more systematic, suggesting it is important to understand how to effectively balance adaptation with being systematic.

In addition to these language design considerations, the \textit{content} of a strategy is likely equally important. For example, Roboto only allows for the targeting of one level of expertise: if a developer does not sufficiently understand how to execute an action or query, they may have no way to proceed and will abandon the strategy.
For expert developers, new strategies may be harder to adopt, as past strategies used frequently over time become habitual, automatic, and unconscious, much as tying a shoe. At the same time, developers who feel they already have an effective strategy may be hard to motivate to change their strategy. In contrast, novices with no strategies or who feel that their strategies are ineffective may be more open to guidance. Alternative strategy description languages might offer affordances for learning unfamiliar concepts and skills referred to in a strategy (e.g., a link to an outside resource explaining a key concept), or for linking strategies with identical approaches but different levels of scaffolding to guide behavior.

Another important question for future research to investigate is whether environments which support developers in executing a strategy are beneficial only when initially learning an unfamiliar strategy or are beneficial each time a developer returns to a strategy. Participants found that the strategy execution environment pushed them to be more systematic, but it is unclear to what degree the change might persist over time if support were removed. In  fields such as medicine and aviation where safety and reliability are paramount, the value of explicit strategies comes in part from ensuring that practitioners are systematic, helping avoid potentially costly errors that may be easily overlooked. In software engineering tasks, the cost of missed steps may vary widely. When debugging a defect, a missed step might prevent a developer from considering a potential location. In other cases, in the moment situated actions based on expertise may lead to better performance. The cost may also vary by task and context, where missed steps might add time to getting to the same ultimate answer or might result in defects with important consequences. There may also be benefits of not being systematic, as developers shortcut long sets of steps through new insights. Understanding the contextual factors influencing the value of being systematic requires a more fine-grained understanding of the moment-to-moment behavior which occurs throughout programming work. 

In our study, we found that guided developers were generally able to understand and follow the strategies. But achieving this required an extensive piloting process, in which we iterated the strategy descriptions to carefully calibrate the level of detail for the specific expertise levels of our participants. In practice, it may be easier to write strategies at multiple levels of detail, supporting participants with varying levels of familiarity of the concepts referenced by strategies. In our current design, each statement is described with progressive disclosure at two levels of detail, with a statement offering a one line high-level description and the comment below clarifying details about what key concepts mean (e.g., defining what ``graphical output'' is in Figure \ref{figure:ToolScreenshot}c). But for complex concepts, it may be helpful to go further, offering longer explanations and examples to effectively teach key concepts. Alternatively, complex queries and actions might themselves be linked to additional sub-strategies and developers given the option to decide that they know the substrategy and can execute it without assistance or to use a strategy tracker to learn it in detail.  

Building a repository of strategies also requires authoring the strategies. As strategies ideally reflect the best known practices, one might imagine software engineering researchers conducting empirical studies to identify various strategies and codifying these explicitly, as we explicitly captured existing strategies in this work. Or developers themselves might play a larger role, as strategy creation is crowdsourced much as sites such as Stack Overflow crowdsource knowledge sharing today. This requires techniques for motivating contribution, collaboratively identifying which strategies are most helpful and when. It may also be helpful to involve users of strategies more directly, collecting feedback to help surface issues and challenges with strategies, confusing or inadequate text, and opportunities for refinement, which others might then address. Companies and organizations might also have a role, as they might wish to establish best practices for their domain or to increase the use of strategies that prioritize specific qualities (e.g., quality over speed, or speed over quality). We envision a rich future ecosystem of strategies, where researchers and developers work together to create, study, and improve strategies over time. 

While our focus has been on explicit strategies for programming, the ideas might also be relevant for knowledge intensive domains outside of programming. In domains where computer and humans both have an important role in solving specific problems, documenting how this collaboration is to take place may have value. Or in more traditional domains where standard operating procedures are commonly used, it may be helpful to offer greater tool support for complex and challenging strategies.

As we noted in the beginning of this paper, programmer productivity is not just a matter of using the right tools, it's also a matter of how they are used. Prior work, and the ideas and evidence in this paper reinforce this view, suggesting that we need substantially more research on strategy languages and tools like Roboto, and research on specific strategies. If we advance our understanding of strategies, not only we we help developers better leverage the tools and skills they have, but in doing so, improve software quality by improving developers' decisions.

\section{Conclusion}
\label{sec:Conclusion}

In this paper, we explored the potential for guiding developers through explicit programming strategies which help to standardize and share steps for solving typical programming problems much as engineering handbooks do in other disciplines. In Roboto, strategies are captured as semi-formal procedures that distribute responsibility to both the developer and computer, where the developer is responsible for reasoning and deciding and the computer helps structure processes and persist information. Using a strategy tracker, developers perform the core reasoning activities, as they are prompted to take action, to gather and process information through queries, and make judgments about their environment through conditions. In turn, the computer helps carry out this reasoning and be more systematic in following the strategy by displaying the current statement, recording variables, and advancing the program counter. In this way, the computer and developer work together. We found that, compared to developers who are free to choose their own strategies, developers given explicit strategies experienced their work as more constrained but also as more organized and systematic. And using explicit strategies enabled developers to be more successful in their work.

Future work should explore many dimensions of explicit programming strategies, including how to support developers in writing them, how to represent strategies support varying levels of expertise, and how to help developers choose strategies appropriate to their tasks. This work could enable a world in which \textit{\textit{how}} to program \textit{well} is not a mysterious, expert skill, hard won only through practice, but something that can be shared widely, for the benefit of all developers.

\begin{acknowledgements}

We thank our study participants for their time. This work was supported in part by the National Science Foundation under grants CCF-1703734 and CCF-1703304.

\end{acknowledgements}





\bibliographystyle{spmpsci}
\bibliography{references}

\begin{thebibliography}{10}
\providecommand{\url}[1]{{#1}}
\providecommand{\urlprefix}{URL }
\expandafter\ifx\csname urlstyle\endcsname\relax
  \providecommand{\doi}[1]{DOI~\discretionary{}{}{}#1}\else
  \providecommand{\doi}{DOI~\discretionary{}{}{}\begingroup
  \urlstyle{rm}\Url}\fi

\bibitem{abbott2015programs}
Abbott, K., Bogart, C., Walkingshaw, E.: Programs for people: What we can learn
  from lab protocols.
\newblock In: Symposium on Visual Languages and Human-Centric Computing
  (VL/HCC), pp. 203--211 (2015)

\bibitem{ames1988achievement}
Ames, C., Archer, J.: Achievement goals in the classroom: Students' learning
  strategies and motivation processes.
\newblock Journal of Educational Psychology \textbf{80}(3), 260 (1988)

\bibitem{anderson1989skill}
Anderson, J.R., Conrad, F.G., Corbett, A.T.: Skill acquisition and the lisp
  tutor.
\newblock Cognitive Science \textbf{13}(4), 467--505 (1989)

\bibitem{Bass2012SAIP}
Bass, L., Clements, P., Kazman, R.: Software Architecture in Practice, 3rd edn.
\newblock Addison-Wesley Professional (2012)

\bibitem{beck2003test}
Beck, K.: Test-driven development: by example.
\newblock Addison-Wesley Professional (2003)

\bibitem{beck2001manifesto}
Beck, K., Beedle, M., Van~Bennekum, A., Cockburn, A., Cunningham, W., Fowler,
  M., Grenning, J., Highsmith, J., Hunt, A., Jeffries, R., et~al.: Manifesto
  for agile software development  (2001)

\bibitem{bielaczyc1995training}
Bielaczyc, K., Pirolli, P.L., Brown, A.L.: Training in self-explanation and
  self-regulation strategies: Investigating the effects of knowledge
  acquisition activities on problem solving.
\newblock Cognition and instruction \textbf{13}(2), 221--252 (1995)

\bibitem{bird2009does}
Bird, C., Nagappan, N., Devanbu, P., Gall, H., Murphy, B.: Does distributed
  development affect software quality? an empirical case study of windows
  vista.
\newblock In: Proceedings of the 31st international conference on software
  engineering, pp. 518--528. IEEE Computer Society (2009)

\bibitem{Bohme2017Debug}
B\"{o}hme, M., Soremekun, E.O., Chattopadhyay, S., Ugherughe, E., Zeller, A.:
  Where is the bug and how is it fixed? an experiment with practitioners.
\newblock In: European Software Engineering Conference and Symposium on the
  Foundations of Software Engineering (ESEC/FSE), pp. 117--128 (2017)

\bibitem{bransford2000people}
Bransford, J.D., Brown, A.L., Cocking, R.R., et~al.: How people learn, vol.~11.
\newblock Washington, DC: National academy press (2000)

\bibitem{ccalicskan2010effects}
{\c{C}}al{\i}{\c{s}}kan, S., Sel{\c{c}}uk, G.S., Erol, M.: Effects of the
  problem solving strategies instruction on the students’ physics problem
  solving performances and strategy usage.
\newblock Procedia-Social and Behavioral Sciences \textbf{2}(2), 2239--2243
  (2010)

\bibitem{chen2002civil}
Chen, W.F., Liew, J.R.: The civil engineering handbook.
\newblock Crc Press (2002)

\bibitem{chi1997quantifying}
Chi, M.T.: Quantifying qualitative analyses of verbal data: A practical guide.
\newblock The Journal of the Learning Sciences \textbf{6}(3), 271--315 (1997)

\bibitem{davies1993models}
Davies, S.P.: Models and theories of programming strategy.
\newblock Int. Journal of Man-Machine Studies \textbf{39}(2), 237--267 (1993)

\bibitem{desoete2001metacognition}
Desoete, A., Roeyers, H., Buysse, A.: Metacognition and mathematical problem
  solving in grade 3.
\newblock Journal of Learning Disabilities \textbf{34}(5), 435--447 (2001)

\bibitem{dieste2017empirical}
Dieste, O., Aranda, A.M., Uyaguari, F., Turhan, B., Tosun, A., Fucci, D., Oivo,
  M., Juristo, N.: Empirical evaluation of the effects of experience on code
  quality and programmer productivity: An exploratory study.
\newblock ESE pp. 1--86 (2017)

\bibitem{einstein1990normal}
Einstein, G.O., McDaniel, M.A.: Normal aging and prospective memory.
\newblock Journal of Experimental Psychology: Learning, Memory, and Cognition
  \textbf{16}(4), 717 (1990)

\bibitem{Falessi2011ArchDecisions}
Falessi, D., Cantone, G., Kazman, R., Kruchten, P.: Decision-making techniques
  for software architecture design: A comparative survey.
\newblock ACM Comput. Surv. \textbf{43}(4), 33:1--33:28 (2011)

\bibitem{falkner2014identifying}
Falkner, K., Vivian, R., Falkner, N.J.: Identifying computer science
  self-regulated learning strategies.
\newblock In: Conference on Innovation \& Technology in Computer Science
  Education, pp. 291--296 (2014)

\bibitem{felleisen2001design}
Felleisen, M.: How to design programs: an introduction to programming and
  computing.
\newblock MIT Press (2001)

\bibitem{francel2001value}
Francel, M.A., Rugaber, S.: The value of slicing while debugging.
\newblock Science of Computer Programming \textbf{40}(2-3), 151--169 (2001)

\bibitem{frick1999formative}
Frick, T., Reigeluth, C.: Formative research: A methodology for creating and
  improving design theories.
\newblock Instructional-design theories. Hillsdale, NJ: Lawrence Erlbaum
  Associates pp. 633--652 (1999)

\bibitem{Gamma:1995}
Gamma, E., Helm, R., Johnson, R., Vlissides, J.: Design Patterns: Elements of
  Reusable Object-oriented Software.
\newblock Addison-Wesley, Boston, MA, USA (1995)

\bibitem{gawande2010checklist}
Gawande, A., Lloyd, J.B.: The checklist manifesto: How to get things right,
  vol. 200.
\newblock Metropolitan Books New York (2010)

\bibitem{gilmore1990expert}
Gilmore, D.: Expert programming knowledge: a strategic approach.
\newblock In: Psychology of programming, pp. 223--234. Academic Press, London
  (1990)

\bibitem{gilmore1991models}
Gilmore, D.J.: Models of debugging.
\newblock Acta psychologica \textbf{78}(1), 151--172 (1991)

\bibitem{haidry2017identifying}
Haidry, S.e.Z., Falkner, K., Szabo, C.: Identifying domain-specific cognitive
  strategies for software engineering.
\newblock In: Conference on Innovation and Technology in Computer Science
  Education (SIGCSE), pp. 206--211 (2017)

\bibitem{hammer2014confusing}
Hammer, D., Berland, L.K.: Confusing claims for data: A critique of common
  practices for presenting qualitative research on learning.
\newblock Journal of the Learning Sciences \textbf{23}(1), 37--46 (2014)

\bibitem{hilton1996appropriateness}
Hilton, J.F.: The appropriateness of the wilcoxon test in ordinal data.
\newblock Statistics in medicine \textbf{15}(6), 631--645 (1996)

\bibitem{horvitz1999principles}
Horvitz, E.: Principles of mixed-initiative user interfaces.
\newblock In: Conference on Human Factors in Computing Systems (CHI), pp.
  159--166 (1999)

\bibitem{Hutchins1995}
Hutchins, E.: Cognition in the Wild.
\newblock MIT Press (1995)

\bibitem{Kerievsky2004Refactoring}
Kerievsky, J.: Refactoring to Patterns.
\newblock Pearson Higher Education (2004)

\bibitem{kersten2006using}
Kersten, M., Murphy, G.C.: Using task context to improve programmer
  productivity.
\newblock In: Foundations of Software Engineering (FSE), pp. 1--11 (2006)

\bibitem{Ko2007InfoNeeds}
Ko, A.J., DeLine, R., Venolia, G.: Information needs in collocated software
  development teams.
\newblock In: ICSE, pp. 344--353 (2007)

\bibitem{ko2015practical}
Ko, A.J., Latoza, T.D., Burnett, M.M.: A practical guide to controlled
  experiments of software engineering tools with human participants.
\newblock ESE \textbf{20}(1), 110--141 (2015)

\bibitem{Ko2019TeachingStrategies}
Ko, A.J., LaToza, T.D., Hull, S., Ko, E.A., Kwok, W., Quichocho, J., Akkaraju,
  H., Pandit, R.: Teaching explicit programming strategies to adolescents.
\newblock In: Symposium on Computer Science Education (SIGCSE), pp. 469--475
  (2019)

\bibitem{ko2010extracting}
Ko, A.J., Myers, B.: Extracting and answering why and why not questions about
  java program output.
\newblock Transactions on Software Engineering and Methodology (TOSEM)
  \textbf{20}(2), 4 (2010)

\bibitem{ko2004designing}
Ko, A.J., Myers, B.A.: Designing the whyline: a debugging interface for asking
  questions about program behavior.
\newblock In: Conference on Human Factors in Computing Systems (CHI), pp.
  151--158 (2004)

\bibitem{ko2005framework}
Ko, A.J., Myers, B.A.: A framework and methodology for studying the causes of
  software errors in programming systems.
\newblock J. of Visual Languages \& Computing \textbf{16}(1-2), 41--84 (2005)

\bibitem{ko2009finding}
Ko, A.J., Myers, B.A.: Finding causes of program output with the java whyline.
\newblock In: Conference on Human Factors in Computing Systems (CHI), pp.
  1569--1578. ACM (2009)

\bibitem{labouvie1976cognitive}
Labouvie-Vief, G., Gonda, J.N.: Cognitive strategy training and intellectual
  performance in the elderly.
\newblock Journal of Gerontology \textbf{31}(3), 327--332 (1976)

\bibitem{LaToza2007Fact}
LaToza, T.D., Garlan, D., Herbsleb, J.D., Myers, B.A.: Program comprehension as
  fact finding.
\newblock In: European Software Engineering Conference and Symposium on the
  Foundations of Software Engineering (ESEC/FSE), pp. 361--370 (2007)

\bibitem{li2015makes}
Li, P.L., Ko, A.J., Zhu, J.: What makes a great software engineer?
\newblock In: ICSE, pp. 700--710 (2015)

\bibitem{locke1984effect}
Locke, E.A., Frederick, E., Lee, C., Bobko, P.: Effect of self-efficacy, goals,
  and task strategies on task performance.
\newblock Journal of applied psychology \textbf{69}(2), 241 (1984)

\bibitem{loksa2016role}
Loksa, D., Ko, A.J.: The role of self-regulation in programming problem solving
  process and success.
\newblock In: Conference on International Computing Education Research (ICER),
  pp. 83--91 (2016)

\bibitem{loksa2016programming}
Loksa, D., Ko, A.J., Jernigan, W., Oleson, A., Mendez, C.J., Burnett, M.M.:
  Programming, problem solving, and self-awareness: effects of explicit
  guidance.
\newblock In: Conference on Human Factors in Computing Systems (CHI), pp.
  1449--1461 (2016)

\bibitem{Fowler1999Refactoring}
Martin~Fowler Kent~Beck, J.B.W.O.D.R.: Refactoring: Improving the Design of
  Existing Code.
\newblock Addison-Wesley Longman Publishing Co., Inc., Boston, MA, USA (1999)

\bibitem{mccartney2007successful}
McCartney, R., Eckerdal, A., Mostrom, J.E., Sanders, K., Zander, C.: Successful
  students' strategies for getting unstuck.
\newblock In: Conference on Innovation and Technology in Computer Science
  Education, pp. 156--160 (2007)

\bibitem{Mernik2005DSL}
Mernik, M., Heering, J., Sloane, A.M.: When and how to develop domain-specific
  languages.
\newblock ACM Computing Surveys \textbf{37}(4), 316--344 (2005)

\bibitem{metzger2004debugging}
Metzger, R.C.: Debugging by thinking: A multidisciplinary approach.
\newblock Digital Press (2004)

\bibitem{MurphySIGCSE08}
Murphy, L., Lewandowski, G., McCauley, R., Simon, B., Thomas, L., Zander, C.:
  Debugging: The good, the bad, and the quirky -- a qualitative analysis of
  novices' strategies.
\newblock In: Symposium on Computer Science Education (SIGCSE), pp. 163--167
  (2008)

\bibitem{pothier2009back}
Pothier, G., Tanter, {\'E}.: Back to the future: Omniscient debugging.
\newblock IEEE Software \textbf{26}(6) (2009)

\bibitem{quinn2011human}
Quinn, A.J., Bederson, B.B.: Human computation: a survey and taxonomy of a
  growing field.
\newblock In: Conference on Human Factors in Computing Systems (CHI), pp.
  1403--1412 (2011)

\bibitem{reason1990}
Reason, J.: Human Error.
\newblock Cambridge University Press (1990).
\newblock \doi{10.1017/CBO9781139062367}

\bibitem{retelny2017no}
Retelny, D., Bernstein, M.S., Valentine, M.A.: No workflow can ever be enough:
  How crowdsourcing workflows constrain complex work.
\newblock Proc. ACM Hum.-Comp. Interact. \textbf{1} (2017)

\bibitem{rist1995program}
Rist, R.S.: Program structure and design.
\newblock Cognitive Science \textbf{19}(4), 507--561 (1995)

\bibitem{robillard2004effective}
Robillard, M.P., Coelho, W., Murphy, G.C.: How effective developers investigate
  source code: an exploratory study.
\newblock Trans. on Software Engineering \textbf{30}(12), 889--903 (2004)

\bibitem{robins2003learning}
Robins, A., Rountree, J., Rountree, N.: Learning and teaching programming: A
  review and discussion.
\newblock Computer Science Education \textbf{13}(2), 137--172 (2003)

\bibitem{roehm2012professional}
Roehm, T., Tiarks, R., Koschke, R., Maalej, W.: How do professional developers
  comprehend software?
\newblock In: International Conference on Software Engineering (ICSE), pp.
  255--265 (2012)

\bibitem{sackman1968exploratory}
Sackman, H., Erikson, W.J., Grant, E.E.: Exploratory experimental studies
  comparing online and offline programming performance.
\newblock Communications of the ACM (CACM) \textbf{11}(1), 3--11 (1968)

\bibitem{salinger2013understanding}
Salinger, S., Prechelt, L.: Understanding Pair Programming: The Base Layer.
\newblock BoD--Books on Demand (2013)

\bibitem{schoenfeld1981episodes}
Schoenfeld, A.H.: Episodes and executive decisions in mathematical problem
  solving.
\newblock In: Annual Meeting of the American Educational Research Association.
  ERIC (1981)

\bibitem{sears2007human}
Sears, A., Jacko, J.A.: The human-computer interaction handbook: fundamentals,
  evolving technologies and emerging applications.
\newblock CRC press (2007)

\bibitem{Shaw:1996}
Shaw, M., Garlan, D.: Software Architecture: Perspectives on an Emerging
  Discipline.
\newblock Prentice-Hall, Upper Saddle River, NJ (1996)

\bibitem{Shull2010TDD}
Shull, F., Melnik, G., Turhan, B., Layman, L., Diep, M., Erdogmus, H.: What do
  we know about test-driven development?
\newblock IEEE Software \textbf{27}(6), 16--19 (2010)

\bibitem{Sillito2008Info}
Sillito, J., Murphy, G.C., De~Volder, K.: Asking and answering questions during
  a programming change task.
\newblock IEEE Trans. Softw. Eng. \textbf{34}(4), 434--451 (2008)

\bibitem{simon1972theories}
Simon, H.A.: Theories of bounded rationality.
\newblock Decision and organization \textbf{1}(1), 161--176 (1972)

\bibitem{suchman1987plans}
Suchman, L.A.: Plans and situated actions: The problem of human-machine
  communication.
\newblock Cambridge university press (1987)

\bibitem{von1995program}
Von~Mayrhauser, A., Vans, A.M.: Program comprehension during software
  maintenance and evolution.
\newblock Computer \textbf{28}(8), 44--55 (1995)

\bibitem{wieringa1998procedure}
Wieringa, D., Moore, C., Barnes, V.: Procedure writing: principles and
  practices.
\newblock IEEE (1998)

\bibitem{xie2018explicit}
Xie, B., Nelson, G.L., Ko, A.J.: An explicit strategy to scaffold novice
  program tracing.
\newblock In: Symposium on Computer Science Education (SIGCSE), pp. 344--349
  (2018)

\bibitem{zamli2001process}
Zamli, K.Z.: Process modeling languages: A literature review.
\newblock Malaysian Journal of Computer Science \textbf{14}(2), 26--37 (2001)

\bibitem{zeller2009programs}
Zeller, A.: Why programs fail: a guide to systematic debugging.
\newblock Elsevier (2009)

\bibitem{ZHANG1994}
Zhang, J., Norman, D.A.: Representations in distributed cognitive tasks.
\newblock Cognitive Science \textbf{18}(1), 87 -- 122 (1994)

\end{thebibliography}

\end{document}